\documentclass[prd,12pt,reprint,nofootinbib,onecolumn,superscriptaddress]{revtex4-2}
\usepackage{bm}
\usepackage{amsmath}
\usepackage{amssymb}
\usepackage{graphicx}
\usepackage{subfigure}
\usepackage{hyperref}
\usepackage{color}
\usepackage{booktabs}
\usepackage{orcidlink}
\usepackage{array}
\usepackage{rotating}
\usepackage{ulem}
\hypersetup{
	colorlinks=true,
	linkcolor=red,
	citecolor=blue,
}

\begin{document}
\title{Detectability to extreme mass ratio inspirals with alternative space-based detector networks}

\author{Chao Zhang\,\orcidlink{0000-0001-8829-1591}}
\email[]{zhangchao1@nbu.edu.cn}
\affiliation{Institute of Fundamental Physics and Quantum Technology, Ningbo University, Ningbo, Zhejiang 315211, China}
\affiliation{Department of Physics, School of Physical Science and Technology, Ningbo University, Ningbo, 315211, China}

\author{Gang Wang\,\orcidlink{0000-0002-9668-8772}}
\email[Corresponding author: ]{gwang@nbu.edu.cn, gwanggw@gmail.com}
\affiliation{Institute of Fundamental Physics and Quantum Technology, Ningbo University, Ningbo, Zhejiang 315211, China}
\affiliation{Department of Physics, School of Physical Science and Technology, Ningbo University, Ningbo, 315211, China}

\date{\today}

\begin{abstract}

Extreme mass ratio inspirals (EMRIs) are among the primary targets for space-based gravitational-wave (GW) detectors, providing valuable opportunities to study stellar-mass compact objects orbiting supermassive black holes (SMBHs) and to probe gravity in the strong-field regime.
As the LISA, TAIJI, and TianQin missions are expected to operate around 2035, joint observations by multiple detectors may provide enhanced measurement capabilities compared to individual missions. 
In this work, we investigate the detectability and parameter-constraint prospects for EMRIs using global detector networks composed of LISA, TianQin, and two alternative TAIJI orbital configurations.
Employing Fisher information matrix analysis, we find that joint observations improve source localization relative to a standalone LISA mission, with sky-localization uncertainties reduced by up to two orders of magnitude for a one-month observation period. We further find that a three-detector network (LISA–TAIJI–TianQin) operating for one month yields parameter constraints comparable to, and in some cases tighter than, those obtained from a one-year observation by LISA alone. 
This result indicates that network observations can partially compensate for shorter observation durations in parameter inference.
Furthermore, we evaluate measurement uncertainties in concurrent dual-signal scenarios: (i) two EMRIs orbiting the same SMBH and (ii) two EMRIs with identical intrinsic parameters but different sky locations and orientations. The results indicate that differences in the detector responses associated with the source geometries reduce correlations between overlapping signals, allowing the sources to remain distinguishable even for a standalone mission. 
The inclusion of multiple detector constellations further improves source localization and tightens parameter constraints in these concurrent-source scenarios.

\end{abstract}

\maketitle

\section{Introduction}

The direct detection of the gravitational wave (GW) event GW150914 by the LIGO and Virgo Collaborations~\cite{Abbott:2016blz,TheLIGOScientific:2016agk} provided direct observational access to gravity in the strong-field regimes. However, ground-based observatories are constrained by the low-frequency cutoff around a few hertz due to seismic and gravity-gradient noise. To access the $10^{-4}$--$10^{-1}$~Hz band rich in astrophysical sources, space-based missions such as the Laser Interferometer Space Antenna (LISA)~\cite{Danzmann:1997hm,Audley:2017drz,LISA:2024hlh}, TianQin~\cite{Luo:2015ght}, and TAIJI~\cite{Hu:2017mde} are planned for the 2030s. LISA comprises three spacecraft in a heliocentric orbit trailing the Earth, forming an equilateral triangle with $2.5 \times 10^6$~km arms to optimize sensitivity in the mHz band. TianQin employs a geocentric configuration with three spacecraft orbiting the Earth, maintaining an equilateral triangle with arm lengths of approximately $1.7 \times 10^5$~km. Its detector plane is oriented toward the reference source RX~J0806.3+1527, yielding focused sensitivity in specific sky regions and shifting its optimal frequency band slightly higher compared to LISA. TAIJI employs a LISA-like heliocentric orbit, comprising three spacecraft that form an equilateral triangle with $3 \times 10^6$~km arms~\cite{Hu:2017mde}.

Recent studies have shown that joint observations using networks composed of LISA, TAIJI, and TianQin can improve measurement capabilities for a variety of astrophysical sources. For localized sources, the complementary antenna patterns and long baselines between detector constellations help reduce degeneracies among source parameters. Consequently, the LISA-TAIJI network can improve the sky localization of massive binary black holes (MBBHs) by up to several orders of magnitude relative to a single detector~\cite{Ruan:2020smc, Wang:2020vkg, Zhang:2020hyx, Zhang:2021wwd, Zhang:2021kkh, Shuman:2021ruh,Bian:2025ifp,Lu:2022wuk}. Similar benefits have also been reported for other source classes, including improved parameter constraints for stellar-mass binary black holes~\cite{Chen:2021sco, Seto:2022xmh} and enhanced resolution of Galactic binaries~\cite{Zhang:2022wcp}. The inclusion of geocentric missions such as TianQin introduces complementary orbital dynamics and distinct spectral sensitivities. Combining heliocentric and geocentric observatories provides additional geometric information that can improve the separation of luminosity distance and orbital inclination~\cite{Huang:2020rjf,10.1093/ptep/ptaa114}. This capability is important for the use of GW sources as standard sirens and may lead to improved constraints on cosmological parameters~\cite{Wang:2019tto, Wang:2021srv, Yang:2021qge, Zhan:2025jqg}. Improved parameter constraints from joint observations can also enhance tests of general relativity and the no-hair theorem~\cite{Wang:2021mou, Zhao:2021zlr, Mu:2025gtg}. Beyond localized sources, global detector networks provide additional capabilities for probing the stochastic gravitational-wave background (SGWB)~\cite{Omiya:2020fvw, Seto:2020mfd, Wang:2021njt, Su:2025nkl, Zhao:2024yau, Bian:2025ifp, Wang:2023ltz}. Studies have shown that varying orbital inclinations and relative phases can improve SGWB sensitivity and facilitate the separation of overlapping signals~\cite{Cai:2023ywp}. In addition, alternative LISA-TAIJI configurations may improve sensitivity to specific polarization modes, enhancing the prospects for detecting parity violation and the quadrupolar nature of the SGWB~\cite{Chen:2024fto, Chen:2024xzw}. From an operational perspective, multi-detector networks can also increase observational redundancy and reduce the impact of temporary interruptions in individual missions.

Extreme mass ratio inspirals (EMRIs), consisting of compact stellar-mass objects inspiraling into massive black holes (MBHs) with mass ratios in the range $10^{-7}$$-$$10^{-4}$, are among the most informative sources for mHz-frequency GW astronomy. Their long-lived and highly relativistic orbital evolution produces complex multi-harmonic GW signals that encode detailed information about the MBH spacetime, including its mass, spin, and multipolar structure, as well as the dynamical environment of galactic nuclei~\cite{Amaro-Seoane:2007osp,Babak:2017tow,Berry:2019wgg,Fan:2020zhy,Zi:2021pdp,Destounis:2020kss,Destounis:2021mqv,Destounis:2021rko,Cardoso:2021wlq}. Consequently, EMRIs provide a valuable probe of both strong-field gravity and MBH populations. To date, studies of EMRI detectability and parameter constraints have primarily focused on individual space-based detectors, with the capabilities of LISA~\cite{Babak:2017tow,Berry:2019wgg}, TianQin~\cite{Fan:2020zhy,Zi:2021pdp}, and TAIJI~\cite{Ren:2023yec} investigated separately. While previous studies have shown that joint detector networks can improve parameter constraints for several classes of GW sources, their impact on EMRI observations, particularly for networks involving LISA, TianQin, and alternative TAIJI orbital configurations, remains largely unexplored. Moreover, realistic observations are expected to contain multiple EMRI signals simultaneously, with detection rates estimated to range from tens to hundreds of events per year~\cite{Amaro-Seoane:2007osp,Babak:2017tow}. These signals may overlap in both time and frequency, introducing source confusion and increasing the complexity of signal separation. The presence of concurrent EMRIs therefore poses additional challenges for source characterization beyond those encountered in isolated-source analyses.

In this work, five dual-detector and two triple-detector networks constructed from LISA, TianQin, and two TAIJI configurations (TAIJIp and TAIJIm, illustrated in Fig.~\ref{fig:LISATIANQINTAIJI}) are investigated to assess the capability of joint observations for characterizing both isolated and concurrent EMRI signals.
The \texttt{FastKerrEccentricEquatorialFlux} model implemented in the \texttt{FEW} package is employed to generate relativistic adiabatic waveforms for eccentric equatorial inspirals~\cite{Chua:2020stf,Katz:2021yft,Speri:2023jte,Chapman-Bird:2025xtd}. The resulting waveforms are processed through a time-domain time-delay interferometry (TDI) operator to incorporate the time-dependent detector response. Parameter uncertainties are evaluated using the Fisher information matrix (FIM) formalism under the assumption of stationary Gaussian noise. We find that joint detector networks improve EMRI parameter constraints, particularly for source localization over short observation periods. For isolated sources, a one-month observation with the detector networks can reduce sky-localization uncertainties by up to two orders of magnitude relative to a standalone LISA mission. In addition, a one-month observation with the three-detector network yields parameter constraints comparable to those obtained from a one-year observation by LISA alone. We further investigate scenarios involving two concurrent EMRI signals and find that differences in the detector responses associated with source geometry reduce correlations between overlapping signals. The inclusion of multiple detector constellations further improves source localization and parameter constraints in these concurrent-source scenarios.

This paper is organized as follows. 
Section~\ref{sec2} outlines the orbital configurations of the space-based detectors--LISA, TianQin, and the two alternative TAIJI orbits--alongside the construction of the joint networks. 
Section~\ref{sec3} introduces the EMRI waveform modeling using the \texttt{FastEMRIWaveforms} package, focusing on the relativistic flux-based framework for eccentric equatorial inspirals into Kerr BHs. 
Section~\ref{sec4} details the implementation of alternative second-generation TDI variables PD4L, as opposed to the standard Michelson combinations, and describes the FIM formalism used to evaluate parameter uncertainties. 
Section~\ref{sec5} presents the numerical results, analyzing the parameter constraint capabilities for both isolated and concurrent EMRIs across different network configurations to quantify the advantages of joint observations. 
Finally, Sec.~\ref {sec6} is devoted to conclusions and discussion.

\section{Alternative space-based GW detector networks}\label{sec2}

\begin{figure}
    \centering
    \includegraphics[width=0.4\textwidth]{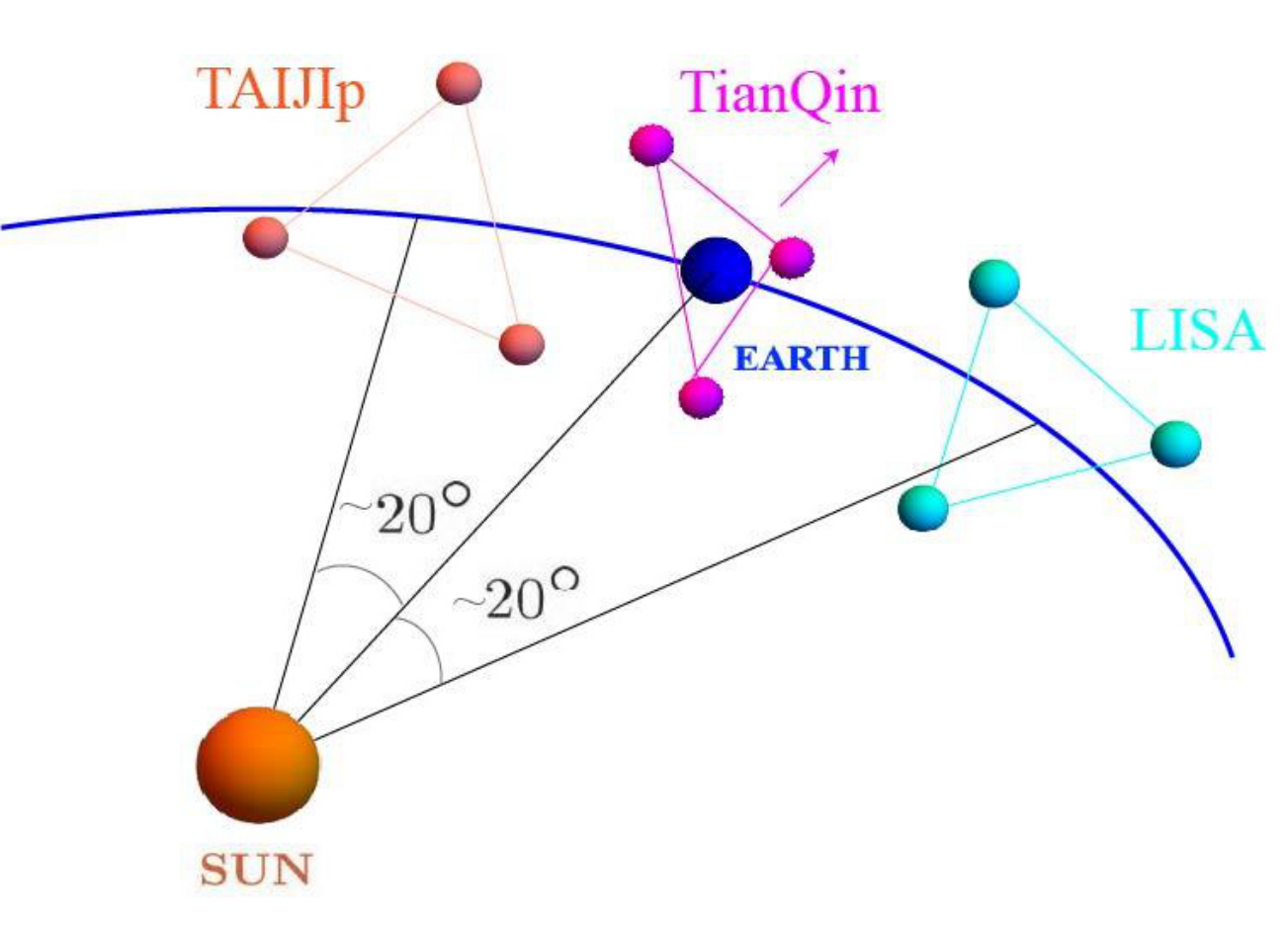}
    \includegraphics[width=0.42\textwidth]{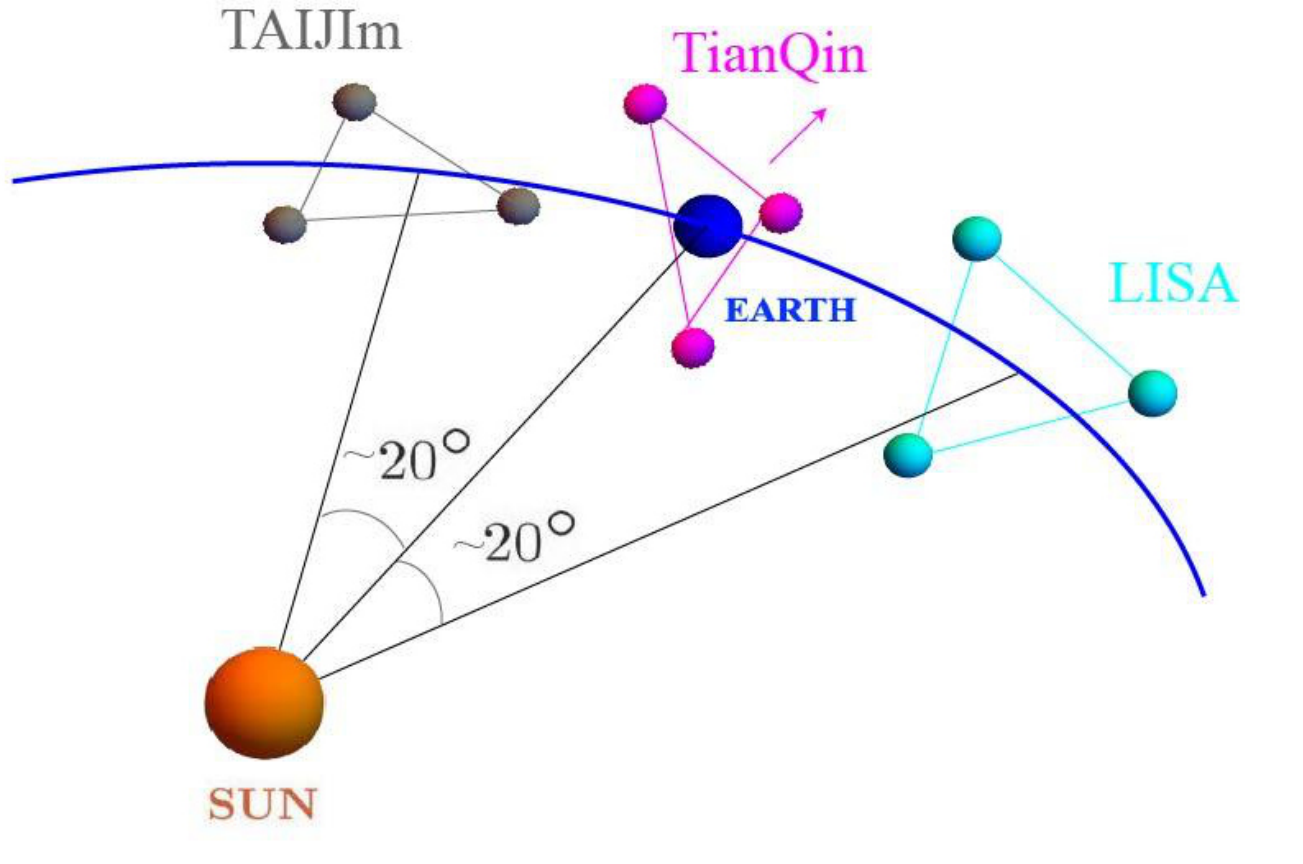}
    \caption{Orbital configurations of the LISA, TianQin, and TAIJI missions. LISA trails the Earth by $\sim 20^\circ$ with a $+60^\circ$ ecliptic inclination. TianQin employs a geocentric orbit pointing toward the reference source RX~J0806.3+1527. TAIJIp leads the Earth by $\sim 20^\circ$ with a $+60^\circ$ inclination (left panel), while TAIJIm leads by $\sim 20^\circ$ with a $-60^\circ$ inclination (right panel).}
    \label{fig:LISATIANQINTAIJI}
\end{figure}

The orbital configurations of the LISA, TAIJI, and TianQin missions are illustrated in Fig.~\ref{fig:LISATIANQINTAIJI}. While sharing overlapping frequency bands, these missions differ in orbital geometries, arm lengths, and sensitivity characteristics, providing complementary observational capabilities for GW detection.

\begin{figure}
	\centering
	\includegraphics[width=0.8\columnwidth]{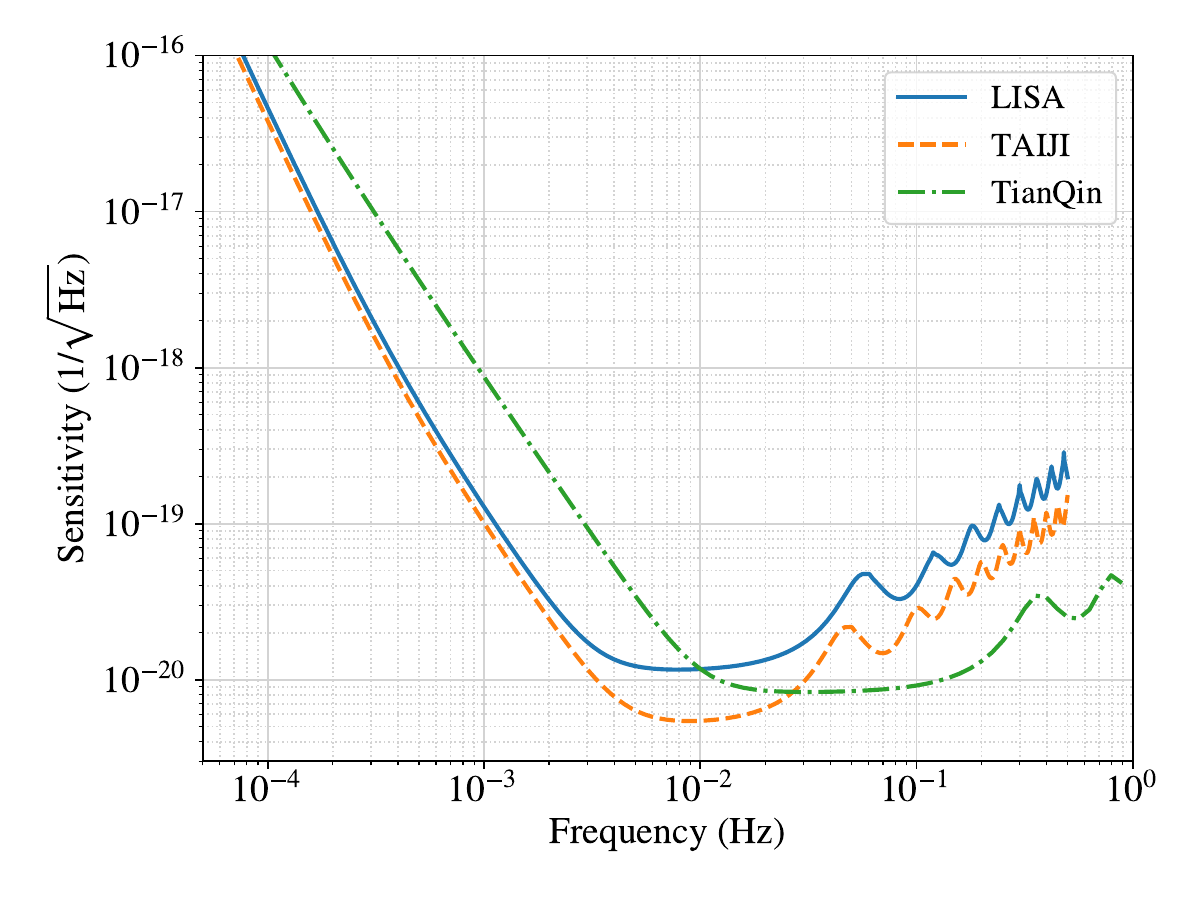}
	\caption{The sky-averaged sensitivity curves for LISA, TianQin, and TAIJI, represented by the (quasi-)orthogonal science channels of their respective TDI configurations.}
\label{fig:sensitivity}
\end{figure}

LISA is currently planned for launch in the mid-2030s~\cite{Audley:2017drz,LISA:2024hlh}. It consists of three spacecraft arranged in an equilateral triangle with arm lengths of $2.5 \times 10^6$~km. 
The constellation orbits the Sun in a heliocentric trailing trajectory, approximately $20^\circ$ behind the Earth, with its orbital plane inclined at $60^\circ$ relative to the ecliptic. The LISA sensitivity band spans the $10^{-4}-10^{-1}$~Hz frequency range, making it well suited for detecting EMRI signals across a broad mass spectrum. 
After the laser frequency noise is suppressed by TDI, the noise sources are dominated by optical metrology system (OMS) noise and test-mass acceleration noise. 
The target upper limit for the OMS noise power spectral density is~\cite{LISA:2024hlh}
\begin{equation}
S_{\text{oms}}^{\text{LISA}}(f) = (15~{\rm pm})^2 \left[1 + \left(\frac{2~{\rm mHz}}{f}\right)^4\right] \text{ Hz}^{-1},
\end{equation}
and the upper bound for the acceleration noise is
\begin{equation}
S_{\text{acc}}^{\text{LISA}}(f) = (3 \times 10^{-15}~{\rm m}/{\rm s}^2)^2 \left[1 + \left(\frac{0.4~{\rm mHz}}{f}\right)^2\right] \left[1 + \left(\frac{f}{8~{\rm mHz}}\right)^4\right] \text{ Hz}^{-1}.
\end{equation}
The sensitivity curve for LISA, assuming the (quasi-)orthogonal science channel of the PD4L TDI scheme (detailed in Section~\ref{sec4}), is shown by the solid blue line in Fig.~\ref{fig:sensitivity}.

The TAIJI mission similarly comprises three spacecraft forming a heliocentric equilateral triangle, but with slightly longer arm lengths of $3 \times 10^6$~km~\cite{Hu:2017mde}. TAIJI could be deployed in an orbit leading/trailing the Earth by approximately $20^\circ$, with its constellation plane similarly inclined relative to the ecliptic. The specific orbital configuration remains flexible, provided the inclination constraint is met. 
To investigate the impact of orbital geometry on network performance, \citet{Wang:2021uih} considered three alternative TAIJI orbits: TAIJIp and TAIJIm lead the Earth by approximately $20^\circ$, with constellation planes inclined at $+60^\circ$ and $-60^\circ$ relative to the ecliptic plane, respectively. This geometry yields a $34.5^\circ$ relative angle between the LISA and TAIJIp constellation planes, and a $71^\circ$ relative angle between the LISA and TAIJIm planes (as illustrated in Fig.~\ref{fig:LISATIANQINTAIJI}). Alternatively, the TAIJIc configuration is colocated and coplanar with LISA. These different geometric orientations relative to LISA lead to distinct antenna-pattern responses and provide a useful framework for assessing network performance.
The OMS noise budget for the TAIJI mission is given by~\cite{Luo:2019zal}
\begin{equation}
S_{\text{oms}}^{\text{TAIJI}}(f) = (8~{\rm pm})^2\left[1+\left(\frac{2~{\rm mHz}}{f}\right)^2\right] \text{ Hz}^{-1},
\end{equation}
and the designed acceleration noise budget is
\begin{equation}
S_{\text{acc}}^{\text{TAIJI}}(f) = (3 \times 10^{-15}~{\rm m}/{\rm s}^2)^2\left[1+\left(\frac{0.4~{\rm mHz}}{f}\right)^2\right]\left[1+\left(\frac{f}{8~{\rm mHz}}\right)^4\right] \text{ Hz}^{-1}.
\end{equation}
The representative sensitivity of TAIJI is depicted by the orange dashed curve in Fig.~\ref{fig:sensitivity}. Owing to its lower OMS noise level, TAIJI is expected to achieve better sensitivity than LISA at frequencies above a few mHz.

The TianQin mission employs a geocentric orbital configuration~\cite{Luo:2015ght}. Three spacecraft form an equilateral triangle with arm lengths of approximately $1.7 \times 10^5$~km, orbiting the Earth while the entire constellation revolves around the Sun synchronously with the Earth. The normal vector of TianQin's detector plane points continuously toward the reference source RX~J0806.3+1527, resulting in enhanced sensitivity toward specific sky regions. The analytical sensitivity model for TianQin is~\cite{Luo:2025ewp, Luo:2015ght,Hu:2018yqb,Lu:2019log}
\begin{equation}
S_n^{\rm {TianQin}}(f) = \frac{10}{3L^2}\left[S_x+\frac{4S_a}{(2\pi f)^4}\left(1+\frac{10^{-4}~{\rm Hz}}{f}\right) \right] \times\left[1+0.6\left(\frac{f}{f_\ast}\right)^2\right],
\end{equation}
where $L\approx 1.7\times 10^8$~m is the nominal arm length, $S_x^{1/2}=1\times 10^{-12}~{\rm m}/{\rm Hz}^{1/2}$ is the displacement measurement noise for each one-way laser link, $S_a^{1/2}=1\times 10^{-15}~{\rm m}/{\rm s}^2/{\rm Hz}^{1/2}$ is the residual acceleration noise for each test mass along the sensitive direction, and $f_\ast=c/(2\pi L)\approx 0.28$~Hz is the transfer frequency. As shown in Fig.~\ref{fig:sensitivity}, TianQin provides better sensitivity than LISA and TAIJI at frequencies above approximately $10^{-2}$~Hz.

In this work, we consider LISA, TianQin, and the two alternative TAIJI orbital configurations (TAIJIp, TAIJIm) to investigate the performance of joint observations for EMRI signals. Seven alternative detector networks are constructed:
\begin{itemize}
    \item LTQ (LISA-TianQin), 
    \item LTp (LISA-TAIJIp), 
    \item LTm (LISA-TAIJIm), 
    \item TpTQ (TAIJIp-TianQin), 
    \item TmTQ (TAIJIm-TianQin)
    \item LTpTQ (LISA-TAIJIp-TianQin)
    \item LTmTQ (LISA-TAIJIm-TianQin)
\end{itemize}
By comparing these network configurations, we quantify the impact of joint observations on parameter constraints, particularly for sky localization, relative to standalone detectors. 
TianQin's geocentric orbit and fixed-source orientation at $(\beta_{\rm TQ}=-4.7^\circ, \lambda_{\rm TQ}=120.5^\circ)$ lead to a seasonal duty cycle. This cycle alternates between three-month scientific observation windows and maintenance periods as the solar illumination angle varies relative to the detector plane. The TianQin operational window is defined as the period during which the angle between the solar vector and the detector plane normal is within $45^\circ$. During these observation windows, the relative angles between the individual constellation planes evolve. The corresponding ranges of relative inclination angles for the dual-detector networks are summarized in Table~\ref{angle}.
\begin{table}[htbp]
    \centering
    \renewcommand{\arraystretch}{1.3}
    \setlength{\tabcolsep}{10pt}
    \caption{\label{angle} Relative orientation angles between the detector plane normals for various dual-detector network configurations. For networks involving TianQin, the indicated ranges account for the angular evolution during its three-month scientific observation window.}
    \begin{tabular}{cc}          
        \toprule
        Network Configuration & Angle between  constellation planes \\
        \midrule
        LTp   & $\sim 34^\circ$     \\
        LTm   & $\sim 71^\circ$     \\
        LTQ   & $42^\circ-71^\circ$ \\
        TpTQ  & $42^\circ-71^\circ$ \\
        TmTQ  & $34^\circ-66^\circ$ \\
        \bottomrule
    \end{tabular}
\end{table}

\section{GW waveform of extreme mass ratio inspirals}\label{sec3}

For EMRIs, the gravitational radiation reaction acting on the stellar-mass compact object comprises the dissipative part and the conservative part~\cite{Cutler:1994pb,Barack:2009ux,Barack:2018yvs, Poisson:2011nh,Mino:1996nk}.
When the timescale of the orbital evolution is much longer than the orbital period, the dynamics can be treated using an adiabatic approximation.
The dissipative part can be calculated from the energy flux radiating at spatial infinity and passing through the horizon of the MBH.
The conservative part, which governs the phase evolution of the waveform, is derived from the gravitational self-force corrections to the orbital motion, contributing to the dephasing of the inspiral over long timescales.
Assuming the motion remains strictly geodesic over several orbital periods, the energy flux from the system can be calculated, yielding the time-averaged rates of change for the orbital parameters. It should be noted that while the energy-flux-driven adiabatic approximation does not incorporate the full gravitational self-force effects---which would provide a more rigorous treatment of the inspiral dynamics---it currently represents the most practical and widely adopted approach for generating EMRI waveforms in large-scale parameter estimation studies. Although more accurate models incorporating self-force effects are anticipated in future developments, the consistent application of the same waveform model across different detector configurations ensures that our conclusions regarding network synergy remain robust.

In Boyer-Lindquist coordinates, the Kerr metric in geometric units is
\begin{equation}
\label{eq:KerrMetric}
\begin{split}
ds^2 =& \left(1-\frac{2r}{\Sigma}\right)dt^2 + \frac{4ar\sin^2\theta}{\Sigma} dt d\varphi 
- \frac{\Sigma}{\Delta} dr^2 - \Sigma d\theta^2 \\
&- \sin^2\theta \left(r^2 + a^2 + \frac{2a^2 r \sin^2\theta}{\Sigma}\right) d\varphi^2,
\end{split}
\end{equation}
where $\Sigma = r^2 + a^2 \cos^2\theta$ and $\Delta = r^2 - 2r + a^2$.
In the Newman-Penrose formalism, the propagating gravitational field can be described by the two complex Newman-Penrose variables
\begin{equation}
\psi_0 = -C_{\alpha\beta\gamma\delta} l^\alpha m^\beta l^\gamma m^\delta, \qquad
\psi_4 = -C_{\alpha\beta\gamma\delta} n^\alpha \bar{m}^\beta n^\gamma \bar{m}^\delta,
\end{equation}
where $C_{\alpha\beta\gamma\delta}$ is the Weyl tensor and the null tetrad is
\begin{equation}
\begin{split}
l^\mu & = \left[(r^2 + a^2)/\Delta, 1, 0, a/\Delta\right], \\
n^\mu & = \left[r^2 + a^2, -\Delta, 0, a\right] / (2\Sigma), \\
m^\mu & = \left[ia\sin\theta, 0, 1, i/\sin\theta\right] / \left(\sqrt{2}(r + ia\cos\theta)\right), \\
\bar{m}^\mu & = \left[-ia\sin\theta, 0, 1, -i/\sin\theta\right] / \left(\sqrt{2}(r - ia\cos\theta)\right).
\end{split}
\end{equation}
The master equation for gravitational tensor ($s=-2$)  perturbations with the source $T$ was derived as \cite{Teukolsky:1973ha}
\begin{widetext}
    \begin{equation}
\label{TB}
\begin{split}
&\left[\frac{(r^2+a^2)^2}{\Delta}-a^2\sin^2{\theta}\right]\frac{\partial^2\psi}{\partial t^2}+\frac{4ar}{\Delta}\frac{\partial^2\psi}{\partial t\partial\varphi}+\left[\frac{a^2}{\Delta}-\frac{1}{\sin^2{\theta}}\right]\frac{\partial^2\psi}{\partial \varphi^2}\\
&\qquad-\Delta^{-s}\frac{\partial}{\partial r}\left(\Delta^{s+1}\frac{\partial\psi}{\partial r}\right)-\frac{1}{\sin\theta}\frac{\partial}{\partial \theta}\left(\sin\theta\frac{\partial\psi}{\partial \theta}\right)-2s\left[\frac{a(r-1)}{\Delta}+\frac{i\cos\theta}{\sin^2{\theta}}\right]\frac{\partial\psi}{\partial \varphi}\\
&\qquad\qquad\qquad\qquad-2s\left[\frac{(r^2-a^2)}{\Delta}-r-ia\cos\theta\right]\frac{\partial\psi}{\partial t}+(s^2\cot^2\theta-s)\psi=4\pi\,\Sigma\, T,
\end{split}
\end{equation}
\end{widetext}
where the explicit field is $\psi=(r-i a\cos\theta)^{4}\psi_4$.
In terms of the eigenfunctions ${_{s}}S_{lm}(\theta)$ \cite{Teukolsky:1973ha,Goldberg:1966uu}, the field $\psi$ and source $T$ can be expanded in the frequency domain as 
\begin{equation}
\psi=\int d\omega \sum_{l,m}R_{\omega lm}(r)~{_{s}}S_{lm}(\theta)e^{-i\omega t+im\varphi},
\end{equation}
\begin{equation}
T_{\omega lm}(r)=\frac{1}{2\pi}\int dt d\Omega ~4\pi \Sigma T ~{_s}S_{lm}(\theta)e^{i\omega t-im\varphi}.
\end{equation}
The radial master function $R_{\omega lm}(r)$ satisfies the inhomogeneous Teukolsky equation
\begin{equation}
\label{Teukolsky}
\Delta^{-s}\frac{d}{d r}\left(\Delta^{s+1}\frac{d R_{\omega lm}}{d r}\right)-V_{T}(r)R_{\omega lm}=T_{\omega lm},
\end{equation}
and the potential function is
\begin{equation}
V_{T}=-\frac{K^2-2is(r-1)K}{\Delta}-4is\omega r+\lambda_{lm\omega},
\end{equation}
 where $K=(r^2+a^2)\omega-am$ and $\lambda_{lm\omega}$ is the corresponding eigenvalue which can be computed by the BH Perturbation Toolkit \cite{BHPToolkit}.
The homogeneous equation \eqref{Teukolsky} admits two linearly independent solutions
\begin{equation}
R_{\omega lm}^{\text{in}} = 
\begin{cases}
B^{\text{tran}} \Delta^{-s} e^{-i\kappa r^*}, & r \to r_+ \\
B^{\text{out}} \dfrac{e^{i\omega r^*}}{r^{2s+1}} + B^{\text{in}} \dfrac{e^{-i\omega r^*}}{r}, & r \to \infty
\end{cases}
\end{equation}
satisfying the asymptotic behavior at the horizon $r_+ = 1 + \sqrt{1-a^2}$ and
\begin{equation}
R_{\omega lm}^{\text{up}} = 
\begin{cases}
D^{\text{out}} e^{i\kappa r^*} + D^{\text{in}} \Delta^{-s} e^{-i\kappa r^*}, & r \to r_+ \\
D^{\text{tran}} \dfrac{e^{i\omega r^*}}{r^{2s+1}}, & r \to \infty
\end{cases}
\end{equation}
satisfying the asymptotic behavior at the infinity,
where $\kappa = \omega - m a/(2r_+)$ and $r^*$ is the tortoise coordinate.
The amplitudes at infinity and the horizon are
\begin{align}
Z_{\omega lm}^{\infty} = & \frac{B^{\text{tran}}}{W} \int_{r_+}^{\infty} \Delta^s R_{\omega lm}^{\text{up}} T_{\omega lm} dr, \\
Z_{\omega lm}^{H} = & \frac{D^{\text{tran}}}{W} \int_{r_+}^{\infty} \Delta^s R_{\omega lm}^{\text{in}} T_{\omega lm} dr,
\end{align}
with the Wronskian $W = \Delta^{s+1}(R^{\text{in}}\partial_r R^{\text{up}} - R^{\text{up}}\partial_r R^{\text{in}}) = 2i\omega B^{\text{in}} D^{\text{tran}}$.
The gravitational energy and angular momentum fluxes radiated to infinity and absorbed by the horizon can be calculated by the amplitudes $Z_{\omega lm}^{\infty}$ and $Z_{\omega lm}^{H}$ through the formula given in Ref. \cite{Teukolsky:1974yv}.
In particular, $\psi_4$ represents outgoing gravitational radiation at infinity and is related to the GW strain by
\begin{equation}
 \lim_{r \to \infty}\psi_4=\frac{1}{2}\left(\ddot{h}_+ - i \ddot{h}_\times\right),
\end{equation}
where $h_+$ and $h_\times$ are the two polarizations of the GW.

Building upon the energy-flux-driven adiabatic framework, the \texttt{FastEMRIWaveforms} (\texttt{FEW}) package provides a highly efficient and modular toolkit for generating EMRI waveforms~\cite{Chua:2020stf,Katz:2021yft,Speri:2023jte,Chapman-Bird:2025xtd}.
The \texttt{FEW} framework leverages precomputed mode data alongside interpolation techniques to rapidly compute adiabatic waveforms, directly addressing the computational bottleneck of tracking multiple harmonic modes over thousands to millions of orbital cycles~\cite{Chapman-Bird:2025xtd}. 
The package supports various waveform models, including fully relativistic flux-based waveforms for eccentric orbits in Schwarzschild spacetime and the Augmented Analytic Kludge model for generic Kerr inspirals~\cite{Chua:2017ujo}.
Both models are optimized for hardware acceleration using graphics processing units, achieving generation times on the order of $\sim 100$~ms~\cite{Chapman-Bird:2025xtd}. 
Recent extensions to the \texttt{FEW} framework have significantly advanced its capabilities, enabling the treatment of eccentric equatorial inspirals into Kerr BHs with spin magnitudes up to $|a| \leq 0.999$, supporting eccentricities $e < 0.9$ and semi-latus recta $p < 200$~\cite{Chapman-Bird:2025xtd}. 
Systematic error characterization demonstrates that this extended model achieves mismatches of $\sim 10^{-5}$ compared to error-free adiabatic waveforms across most of the parameter space relevant to LISA~\cite{Chapman-Bird:2025xtd,Khalvati:2024tzz}. 
In this work, we generate EMRI waveforms utilizing the \texttt{FEW} package.
Specifically, we employ the \texttt{FastKerrEccentricEquatorialFlux} model in the source frame, which integrates a flux-based trajectory module, an interpolation-based amplitude module, and a mode-summation module to produce precise adiabatic waveforms for eccentric equatorial inspirals into Kerr BHs~\cite{Chapman-Bird:2025xtd}.
These generated waveforms are subsequently embedded into our TDI framework for FIM analysis under the various space-based detector network configurations.

\section{Time-Delay Interferometry and Fisher Information Matrix Formalism} \label{sec4}

TDI is a fundamental technique employed by space-based GW detectors to suppress the otherwise overwhelming laser frequency noise arising from unequal and time-varying interferometer arm lengths~\cite{1997SPIE.3116..105N,1999ApJ...527..814A,Tinto:2003vj,Shaddock:2003dj}. First-generation TDI schemes are sufficient for canceling laser frequency noise under static unequal-arm conditions~\cite{Estabrook:2000ef,Tinto:2001ii,Tinto:2001ui,Hogan:2001jn,Armstrong:2001uh,Prince:2002hp,Tinto:2002de,Shaddock:2003dj,Tinto:2003uk,Nayak:2003na,Tinto:2004nz,Romano:2006rj,Zhang:2020khm}, whereas second-generation TDI extends this capability to scenarios involving time-varying arm lengths, providing more robust noise reduction~\cite{Tinto:2003vj,Cornish:2003tz,Vallisneri:2004bn,Krolak:2004xp,Vallisneri:2005ji,Wang:2017aqq,Wang:2020fwa,RajeshNayak:2004jzp,Nayak:2005un}. Among the various second-generation TDI configurations, different schemes exhibit distinct characteristics, including differences in effective delay spans, transfer-function structures, and null-frequency distributions.

In this work, we adopt the PD4L configuration, a second-generation TDI scheme with a reduced effective delay span. It is interpreted as a hybrid construction combining the first-generation Beacon and Monitor observables~\cite{Wang:2011tlj}. The PD4L scheme belongs to the class of 16-link second-generation TDI combinations and satisfies the laser-noise cancellation requirements under dynamically unequal-arm conditions. The explicit spacecraft routing sequences for the three ordinary channels of PD4L are defined as follows:
\begin{align}
\text{PD4L-1} &: \overrightarrow{1 2 3 2} \ \overleftarrow{2 1 2} \ \overrightarrow{2 3 2 1} \ \overleftarrow{1 3 2 3} \ \overrightarrow{3 1 3} \ \overleftarrow{3 2 3 1}\,, \\
\text{PD4L-2} &: \overrightarrow{2 3 1 3} \ \overleftarrow{3 2 3} \ \overrightarrow{3 1 3 2} \ \overleftarrow{2 1 3 1} \ \overrightarrow{1 2 1} \ \overleftarrow{1 3 1 2}\,, \\
\text{PD4L-3} &: \overrightarrow{3 1 2 1} \ \overleftarrow{1 3 1} \ \overrightarrow{1 2 1 3} \ \overleftarrow{3 2 1 2} \ \overrightarrow{2 3 2} \ \overleftarrow{2 1 2 3}\,.
\end{align}
The three ordinary PD4L channels are constructed through cyclic permutations of the spacecraft indices. This sequence formulation adopts the notational convention established in Ref.~\cite{Vallisneri:2005ji}. In this framework, the directional arrows dictate the temporal progression: a rightward arrow ($\rightarrow$) specifies forward time evolution from left to right, whereas a leftward arrow ($\leftarrow$) indicates backward temporal flow. The integer digits explicitly denote the individual spacecraft nodes. 
The corresponding quasi-orthogonal observables (A,E,T) are then obtained via the orthogonal transformation \cite{Prince:2002hp,Vallisneri:2007xa}:
\begin{equation} \label{eq:abc2AET}
\begin{bmatrix}
\mathrm{A}  \\ \mathrm{E}  \\ \mathrm{T}
\end{bmatrix}
 =
\begin{bmatrix}
-\frac{1}{\sqrt{2}} & 0 & \frac{1}{\sqrt{2}} \\
\frac{1}{\sqrt{6}} & -\frac{2}{\sqrt{6}} & \frac{1}{\sqrt{6}} \\
\frac{1}{\sqrt{3}} & \frac{1}{\sqrt{3}} & \frac{1}{\sqrt{3}}
\end{bmatrix}
\begin{bmatrix}
\mathrm{PD4L1} \\ \mathrm{PD4L2}  \\ \mathrm{PD4L3}
\end{bmatrix}.
\end{equation}

Compared with conventional second-generation TDI Michelson observables, PD4L is characterized by a reduced effective delay span. While Michelson observables typically involve an effective delay footprint of $8L$ (where $L$ denotes the nominal light-travel time of a single arm), PD4L reduces this span to $4L$. Additionally, PD4L exhibits fewer and less densely distributed transfer-function null frequencies in the TDI channels. These properties result in smaller boundary losses for finite-duration data segments, reduced frequency aliasing for rapidly evolving signals, and smoother transfer functions with more stable null-channel behavior. Such characteristics can be advantageous for GW data analysis involving rapidly evolving signals or interrupted observations~\cite{Wang:2025mee,Wang:2025voa}.
These advantages become highly relevant for signals whose frequencies evolve significantly over the TDI delay time, such as EMRIs entering the high-frequency regime. In such scenarios, the standard factorized frequency-domain approximation may partially break down due to the finite temporal footprint of the TDI operator. 
Shortened-span configurations can alleviate these effects and improve the accuracy of the detector-response modeling. To circumvent potential inaccuracies in frequency-domain response modeling, TDI is applied directly to the time-domain EMRI waveforms in this work.

For an EMRI signal, the time-domain TDI response can be expressed as
\begin{equation}
    h_\mathrm{TDI} (t, \xi) = \mathcal{T} h (t, \xi)
\end{equation}
where $\mathcal{T}$ represents the TDI operator, and the parameter vector $\boldsymbol{\xi}$ comprises twelve parameters:
\begin{equation}
\xi=(m_1, m_2, a, p_0, e_0, \lambda, \beta, \iota, \psi, \ln d_L, \Phi_{r0}, \Phi_{\phi0}).
\end{equation}
Here, $m_1$ and $a$ are the mass and dimensionless spin of the MBH, $m_2$ is the mass of the secondary compact object, $p_0$ and $e_0$ are the initial orbital semi-latus rectum and eccentricity, $\lambda$ and $\beta$ are the ecliptic longitude and latitude, $d_L$ is the luminosity distance, $\iota$ is the orbital inclination angle, $\psi$ is the polarization angle, and $\Phi_{r0}$ and $\Phi_{\phi0}$ are the initial radial and azimuthal phases. The source-frame waveforms are generated using \texttt{FEW}~\cite{Chua:2020stf,Katz:2021yft}, and the TDI operations are implemented using \texttt{SATDI}~\cite{Wang:2024ssp}.

The matched-filter signal-to-noise ratio (SNR) of a GW signal in a given detector is evaluated by summing the contributions from its three quasi-orthogonal TDI channels:
\begin{equation}
\rm{SNR}=\sqrt{ \sum_{\rm TDI=A, E, T} \left\langle \tilde{h}_\mathrm{TDI} | \tilde{h}_\mathrm{TDI} \right\rangle},
\end{equation}
where $\tilde{h}_\mathrm{TDI}$ is the frequency-domain strain obtained via the Fourier transform of the time-domain TDI response. The noise-weighted inner product between two frequency-domain signals $\tilde{h}_1$ and $\tilde{h}_2$ is defined as
\begin{equation}\label{product}
\left\langle \tilde{h}_{1} | \tilde{h}_{2}\right\rangle=4 \Re \int_{f_{\min }}^{f_{\max }} \frac{\tilde{h}_{1}(f) \tilde{h}_{2}^{*}(f)}{S_{n}(f)} df,
\end{equation}
where $S_n(f)$ is the noise PSD of the respective TDI channel. The integration limits are set to $f_{\text{min}}=10^{-4}$~Hz and $f_{\text{max}}=0.1$~Hz. This upper frequency cutoff is set to 0.1 Hz for two practical reasons. First, the EMRIs considered in this study accumulate negligible SNR above $0.1$~Hz; second, setting the corresponding time-domain sampling frequency to $0.2$~Hz significantly reduces the computational cost of applying the TDI operator.

Under the assumptions of stationary Gaussian noise and sufficiently high SNR, the posterior distribution of the source parameters can be approximated by a multivariate Gaussian centered on the true parameter values. Assuming flat priors, the covariance matrix is given by the inverse of the FIM, whose elements for a single detector are computed as
\begin{equation}
\Gamma_{i j,\mathrm{detector}}= \sum_{\rm TDI=A, E, T} \left\langle\left.\frac{\partial \tilde{h}_\mathrm{TDI}}{\partial \xi_{i}}\right| \frac{\partial \tilde{h}_\mathrm{TDI}}{\partial \xi_{j}}\right\rangle_{\xi=\hat{\xi}}.
\end{equation}
The statistical $1\sigma$ uncertainty $\Delta \xi_i$ on the parameter $\xi_i$ and the correlation coefficients $c_{ij}$ between parameters $\xi_i$ and $\xi_j$ are derived from the covariance matrix $\boldsymbol{\Sigma}_{\mathrm{detector}} = \boldsymbol{\Gamma}_{\mathrm{detector}}^{-1}$ as
\begin{equation}
\Delta \xi_i = \sqrt{\Sigma_{ii,\mathrm{detector}}}, \qquad c_{ij} = \frac{\Sigma_{ij,\mathrm{detector}}}{\Delta \xi_i \Delta \xi_j}.
\end{equation}
Furthermore, the angular uncertainty for sky localization is evaluated as~\cite{Cutler:1997ta}
\begin{equation}
\Delta \Omega = 2\pi |\cos\beta| \sqrt{\Sigma_{\lambda\lambda,\mathrm{detector}} \Sigma_{\beta\beta,\mathrm{detector}} - \Sigma_{\lambda\beta,\mathrm{detector}}^2}.
\end{equation}
For a detector network with statistically independent noise, the total FIM is obtained by summing the Fisher matrices of the individual detectors~\cite{Cutler:1997ta}. The global covariance matrix is then obtained by inverting this combined matrix:
\begin{equation}
    \bf \Sigma_\mathrm{joint}=(\Gamma_{\mathrm{detector1}}+\Gamma_{\mathrm{detector2}}+\cdots)^{-1}.
\end{equation}
The corresponding parameter uncertainties and correlation coefficients are then obtained from the elements of $\boldsymbol{\Sigma}_{\mathrm{joint}}$.

\section{Detectability of EMRIs with detector networks} \label{sec5}

In this section, we assess the EMRI detection capabilities of the various space-based detector networks under two distinct scenarios. 
The first scenario investigates the parameter uncertainties for an isolated EMRI source. Here, we compare the performance of dual- and triple-detector networks over a one-month observation period against a one-year baseline from the standalone LISA mission. This comparison is intended to quantify the impact of joint observations on parameter constraints and to assess how network observations compare with a one-year standalone LISA observation.
The second scenario extends our analysis to the case of two concurrent EMRIs---a situation that becomes increasingly probable given the expected detection rates and the long operational lifespans of space-borne observatories. This investigation evaluates the impact of signal overlap on overall detectability and parameter degeneracies.

\subsection{Parameter constraints for an isolated EMRI}

For the analysis of isolated EMRIs, a representative source configuration is adopted, featuring a primary MBH of mass $m_1=10^6~M_{\odot}$ and a secondary compact object of $m_2=10~M_{\odot}$. The primary component is characterized by a dimensionless spin of $a=0.9$, while the spin of the secondary is neglected. The system is initialized with an eccentricity of $e_0=0.1$ and placed at a fiducial luminosity distance of $d_L=1~{\rm Gpc}$. To sample the variation in detector responses across different source locations and orientations, 1000 independent realizations are generated by randomly drawing the extrinsic parameters.
Assuming an isotropic and unpolarized source population, the polarization angle $\psi$ is uniformly sampled in the range $[0, \pi)$, the initial radial and azimuthal phases $\Phi_{r0}$ and $\Phi_{\phi0}$ are drawn from uniform distributions over the interval $[0, 2\pi)$, the ecliptic latitude $\beta$ and the inclination angle $\iota$ are sampled such that $\sin\beta$ and $\cos\iota$ are uniformly distributed over $[-1, 1]$, the ecliptic longitude $\lambda$ is uniformly sampled in the range $[-\pi, \pi)$. 
The initial semi-latus rectum $p_0$ is iteratively adjusted to ensure the specified observation duration prior to the final plunge.

\begin{figure*}[htbp]
    \centering
    \includegraphics[width=0.95\textwidth]{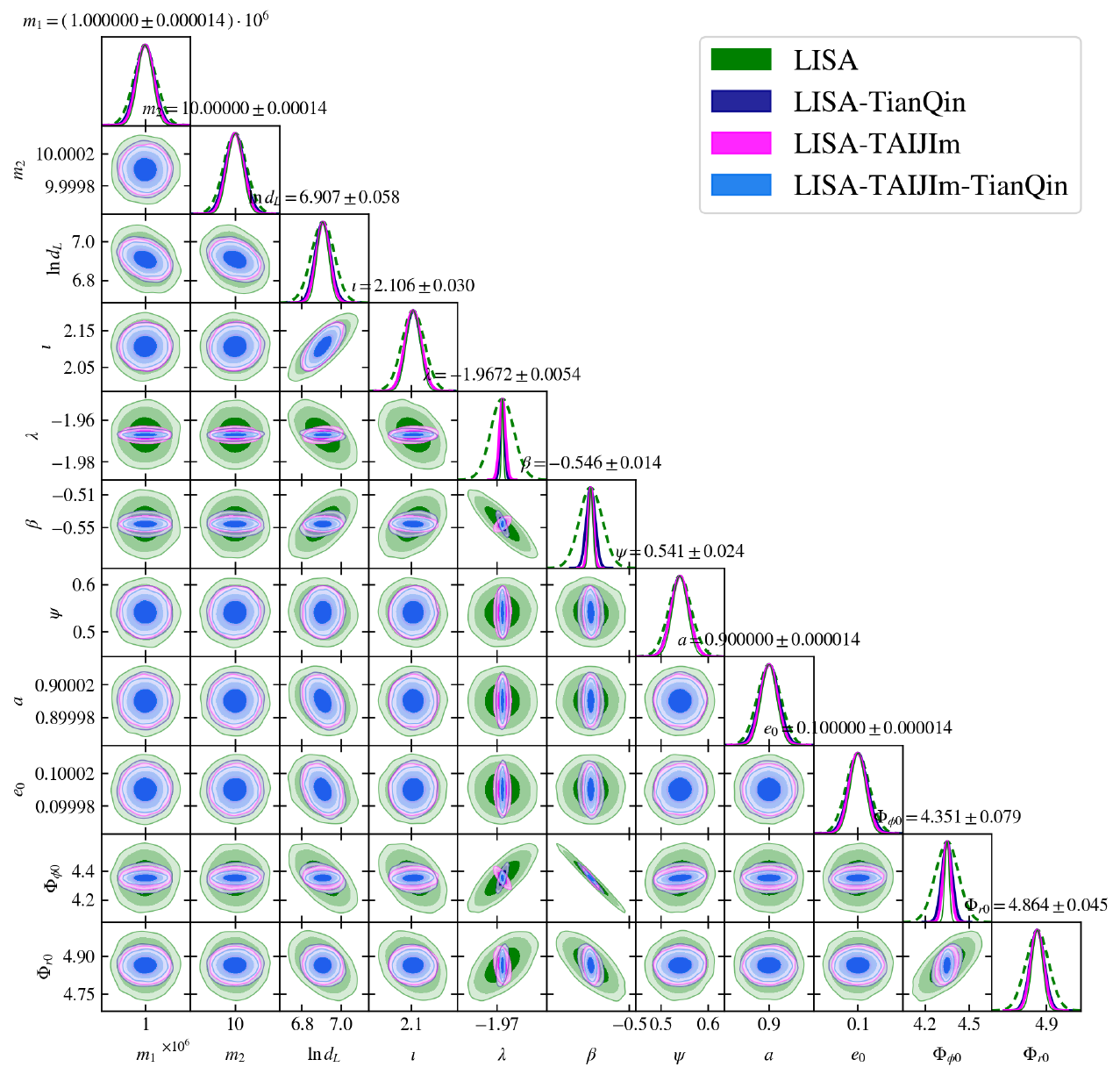}
    \caption{Corner plot showing the Fisher-matrix forecast of the parameter uncertainties for a representative EMRI source. The diagonal panels display the marginalized one-dimensional Gaussian distributions, while the off-diagonal panels show the corresponding two-dimensional confidence regions. The values above each column indicate the $1\sigma$ parameter uncertainties obtained from the standalone LISA mission. Contours denote the $1\sigma$, $2\sigma$, and $3\sigma$ confidence regions.}
    \label{fig:corner1}
\end{figure*}

As a representative example, we consider an EMRI source parameterized by its intrinsic properties: $m_1 = 10^6~M_\odot$, $m_2 = 10~M_\odot$, $a = 0.9$, and $e_0 = 0.1$. 
The extrinsic and phase parameters, which govern its spatial position, orientation, and initial orbital state, are configured as $\iota = 2.10$, $\lambda = -1.97$, $\beta = -0.546$, $\psi = 0.54$, $\Phi_{\phi 0} = 4.35$, and $\Phi_{r0} = 4.86$. Assuming these true values as the distribution centers, the resulting Fisher-matrix covariance yields the corner plot presented in Fig.~\ref{fig:corner1}. 
Three representative joint observation networks are evaluated and compared against the standalone LISA mission (indicated by the green contours). 
The dark blue contours denote the parameter constraints derived from the LISA-TianQin network, the magenta contours correspond to the LISA-TAIJIm network, and the light blue contours correspond to the constraints obtained with the triple-detector network.
The numerical values at the top of each column indicate the $1\sigma$ measurement uncertainties obtained using the single LISA mission. 
As indicated by the covariance contours, the impact of network observations is most evident in the sky localization of the source (tightly constraining $\lambda$ and $\beta$), whereas the constraints on intrinsic parameters show comparatively smaller improvements.

\begin{figure*}[htbp]
	\centering
	\includegraphics[width=0.80\textwidth]{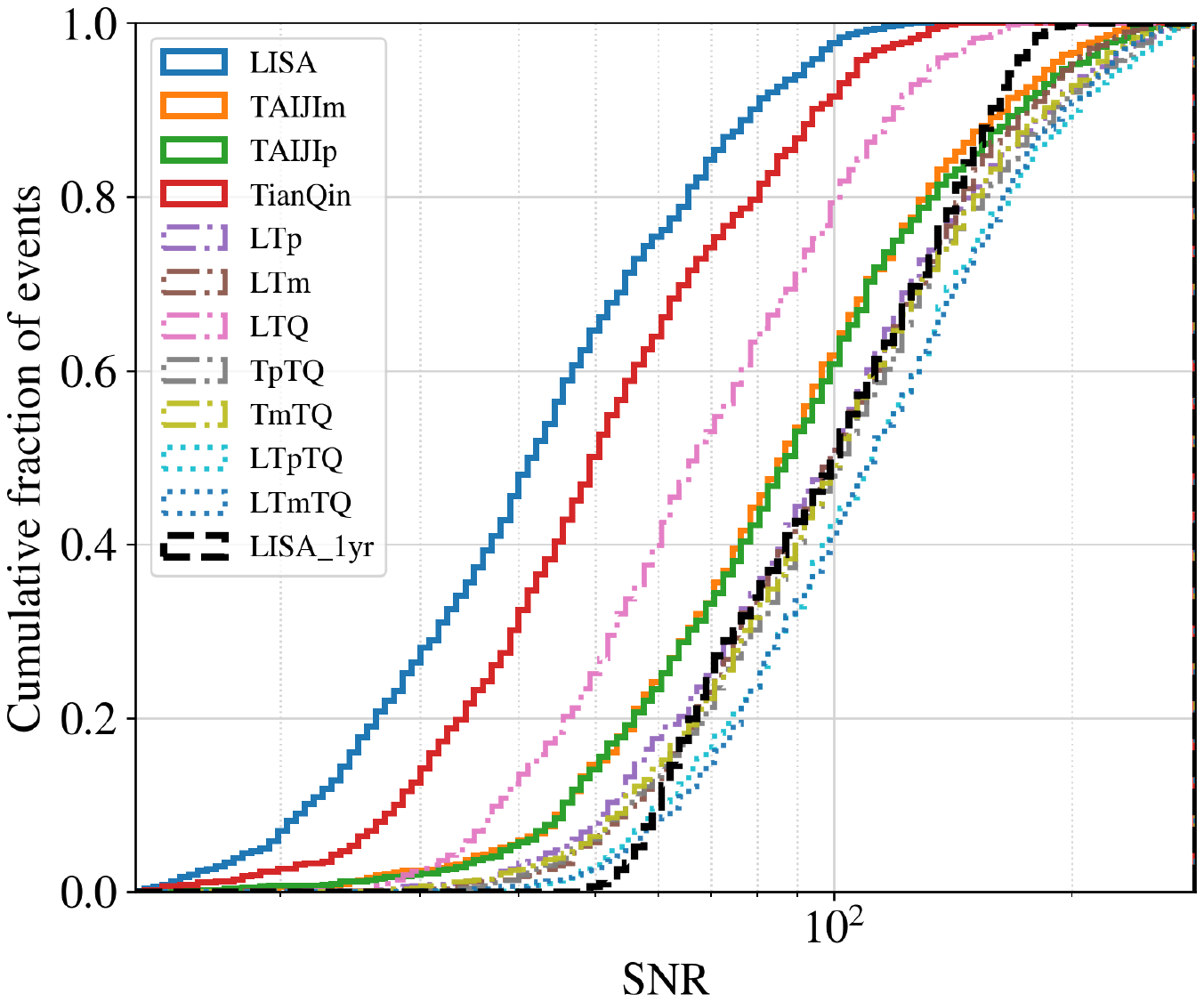}
    \caption{Cumulative distributions of the optimal SNR for 1000 simulated isolated EMRI events under different detector configurations. Solid curves denote standalone missions (LISA, TianQin, TAIJIp, and TAIJIm) with a one-month observation period. Dash-dotted and dotted curves correspond to dual- and triple-detector networks, respectively, over the same observation duration. The thick black dashed curve shows the SNR distribution obtained from a one-year standalone LISA observation. The rightward shift of the network distributions illustrates the substantial increase in accumulated SNR provided by joint observations.}
    \label{fig:SNRhist}
\end{figure*}

Figure~\ref{fig:SNRhist} illustrates the cumulative distribution of the optimal SNR for the 1000 simulated isolated EMRI events under different detector configurations, while the corresponding median values are summarized in the first row of Table~\ref{tab:SEMRItable}. For a one-month observation, the standalone LISA mission has the lowest SNR distribution (solid blue curve), with a median optimal SNR of $41.8$. TianQin (solid red curve) and the two TAIJI configurations (solid orange and green curves) achieve higher SNRs, consistent with their sensitivity curves shown in Fig.~\ref{fig:sensitivity}. When these observatories are combined into detector networks, the cumulative distributions shift toward larger SNRs, reflecting the combined signal contributions from multiple detectors. In particular, the heliocentric two-detector networks (LTp and LTm) increase the median SNR to approximately $100$, while the inclusion of TianQin increases the median SNR to $\sim114$ for the three-detector networks (LTpTQ and LTmTQ). Several one-month network configurations achieve SNRs comparable to, or larger than, those obtained from a one-year standalone LISA observation (black dashed curve).

\begin{figure*}[htbp]
	\centering
	\includegraphics[width=0.90\textwidth]{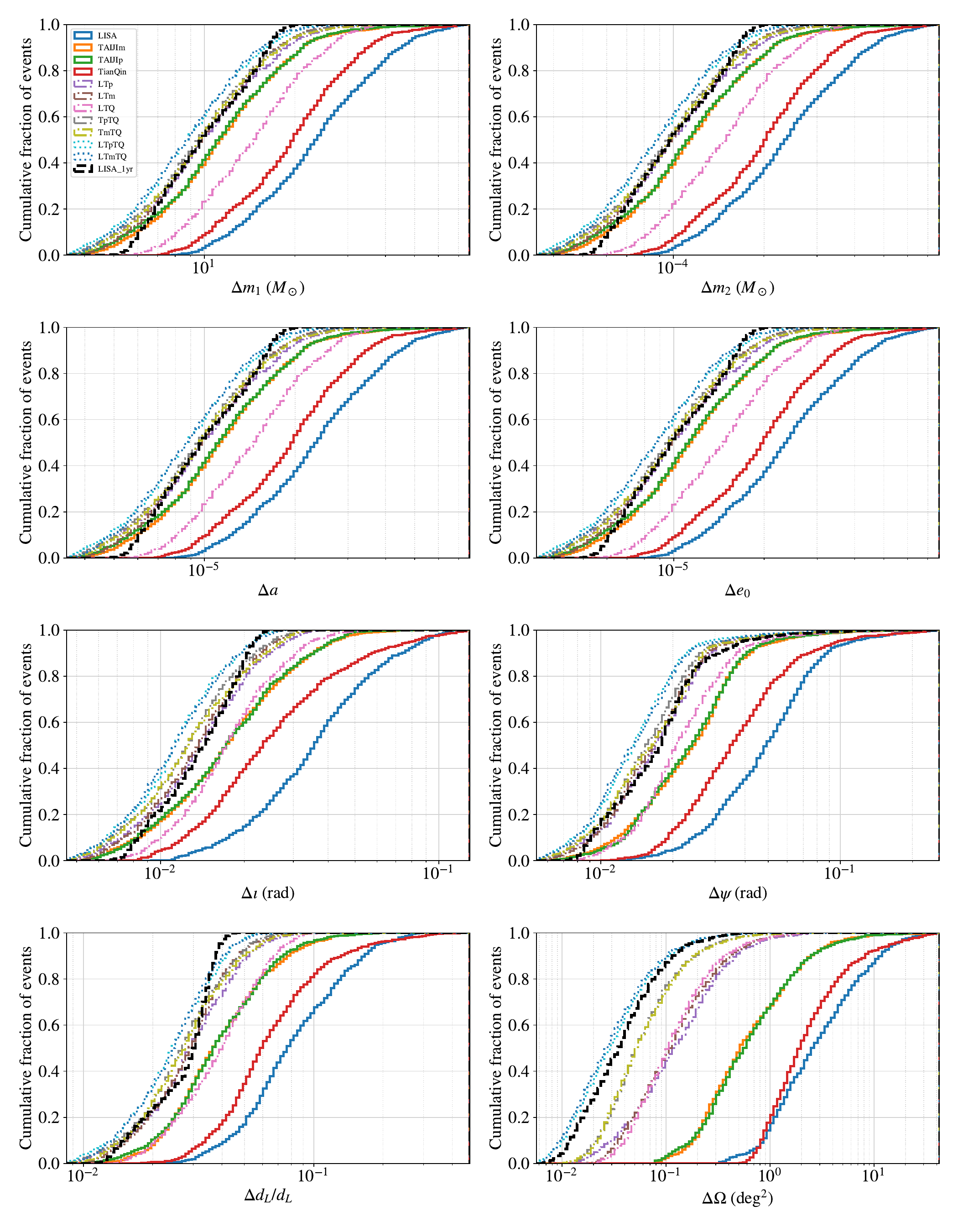}
    \caption{Cumulative distributions of the $1\sigma$ parameter uncertainties for 1000 simulated isolated EMRI events. The panels show the estimation accuracy for intrinsic parameters ($m_1$, $m_2$, $a$, $e_0$) and selected extrinsic parameters ($\iota$, $\psi$, $d_L$, and $\Delta\Omega$). Solid curves represent standalone missions with a one-month observation period, while dash-dotted and dotted curves correspond to dual- and triple-detector networks, respectively. The thick black dashed curve indicates the reference performance of a one-year standalone LISA observation. The leftward shift of the network distributions demonstrates improved parameter-estimation accuracy, particularly for sky localization.}
	\label{fig:SEMRIhist}
\end{figure*}

The impact of the SNR enhancement on parameter constraint is illustrated in Fig.~\ref{fig:SEMRIhist}, while the corresponding median uncertainties are listed in Table~\ref{tab:SEMRItable}. For most intrinsic parameters, including the primary and secondary masses ($m_1$, $m_2$), the primary BH spin ($a$), and the initial eccentricity ($e_0$), the improvement closely follows the expected FIM scaling $\propto 1/\mathrm{SNR}$. For example, the median uncertainty of the primary BH mass decreases from $23.9\,M_\odot$ for one-month LISA observations to $8.80\,M_\odot$ for the three-detector networks, consistent with the corresponding increase in SNR by a factor of 2.7. This behavior suggests that the improvement for these parameters is primarily associated with the increased SNR, rather than with the reduction of additional parameter degeneracies.
A similar behavior is observed for several extrinsic parameters, including the luminosity distance $d_L$, inclination angle $\iota$, and polarization angle $\psi$. Because our analysis employs the \texttt{FEW} waveform framework, which naturally incorporates higher-order harmonics, the distance--inclination degeneracy is already largely broken at the waveform-model level~\cite{Zhang:2023ceh}. Consequently, improvements in these parameters are largely consistent with the increase in SNR provided by the detector networks.

A different trend is observed for sky localization. Unlike the intrinsic parameters, the sky-position uncertainty $\Delta\Omega$ depends not only on the increase in SNR but also on the geometric configuration of the detector network, as shown in bottom right panel of Fig. \ref{fig:SEMRIhist}. 
For a one-month observation, standalone LISA achieves a median localization uncertainty of $2.67~\mathrm{deg}^2$.
The two-detector heliocentric networks improve this precision by more than an order of magnitude, reaching $0.11~\mathrm{deg}^2$ for LTm, while the three-detector configurations further reduce the uncertainty to approximately $0.029~\mathrm{deg}^2$. The reduction in sky-localization uncertainty is stronger than would be expected from SNR scaling alone. The behavior can be attributed to the large spatial separations and distinct antenna pattern of the individual constellations. GW arrive at the detectors with measurable time delays and phase differences, providing additional geometric information that improves source localization of the source. As a result, the detector network helps reduce the sky-position degeneracies that remain unresolved in a single-detector observation.

\begin{widetext}
\begin{sidewaystable*}
\centering
\caption{Median values of the optimal SNR and the corresponding $1\sigma$ parameter uncertainties for the simulated isolated EMRI population. The quoted values represent the medians of 1000 random source realizations, while superscripts and subscripts denote the 84th and 16th percentiles, respectively. Columns from LISA to LTmTQ correspond to one-month observations, whereas the final column (LISA-1yr) provides the reference results for a one-year standalone LISA observation.}
\setlength{\tabcolsep}{3pt}
\renewcommand{\arraystretch}{1.5}
\begin{tabular}{l|cccc|ccccc|cc|c}
\hline
\textbf{Detector} & LISA & TAIJIm & TAIJIp & TQ & LTp & LTm & LTQ & TpTQ & TmTQ & LTpTQ & LTmTQ & LISA-1yr \\
\hline
$\rm{SNR}$ & $41.8^{+28.3}_{-17.3}$ & $87.3^{+54.6}_{-34.6}$ & $89.1^{+60.9}_{-37.0}$ & $50.3^{+35.7}_{-18.9}$ & $101^{+62.9}_{-42.6}$ & $100^{+56.2}_{-36.1}$ & $67.5^{+42.0}_{-24.1}$ & $104^{+66.2}_{-39.6}$ & $103^{+62.1}_{-40.0}$ & $114^{+68.1}_{-43.7}$ & $114^{+63.5}_{-41.4}$ & $101^{+49.1}_{-36.5}$ \\
$\Delta m_1~(M_{\odot})$ & $23.9^{+16.9}_{-9.68}$ & $11.5^{+7.51}_{-4.40}$ & $11.2^{+7.99}_{-4.56}$ & $20.0^{+12.0}_{-8.28}$ & $9.90^{+7.22}_{-3.80}$ & $10.0^{+5.66}_{-3.60}$ & $14.8^{+8.26}_{-5.67}$ & $9.60^{+5.88}_{-3.73}$ & $9.73^{+6.18}_{-3.67}$ & $8.78^{+5.45}_{-3.28}$ & $8.80^{+5.03}_{-3.15}$ & $9.92^{+5.62}_{-3.22}$ \\
$\Delta m_2 \ (10^{-5}~M_{\odot})$ & $24.1^{+16.9}_{-9.72}$ & $11.5^{+7.53}_{-4.40}$ & $11.3^{+8.01}_{-4.59}$ & $20.5^{+12.3}_{-8.40}$ & $9.94^{+7.21}_{-3.81}$ & $10.0^{+5.64}_{-3.62}$ & $15.0^{+8.38}_{-5.65}$ & $9.66^{+5.92}_{-3.74}$ & $9.81^{+6.17}_{-3.69}$ & $8.84^{+5.44}_{-3.30}$ & $8.85^{+5.02}_{-3.16}$ & $9.91^{+5.61}_{-3.21}$ \\
$\Delta d_L/d_L~(10^{-2})$ & $7.81^{+6.46}_{-2.87}$ & $3.71^{+3.01}_{-1.32}$ & $3.73^{+2.77}_{-1.38}$ & $6.09^{+5.28}_{-2.11}$ & $2.94^{+1.74}_{-1.13}$ & $2.95^{+1.33}_{-1.13}$ & $4.05^{+2.19}_{-1.65}$ & $2.73^{+1.65}_{-1.05}$ & $2.78^{+1.68}_{-1.05}$ & $2.45^{+1.43}_{-0.96}$ & $2.48^{+1.22}_{-0.98}$ & $3.06^{+0.62}_{-1.35}$ \\
$\Delta \iota ~(10^{-2}~\rm{rad})$ & $3.61^{+2.60}_{-1.51}$ & $1.75^{+1.32}_{-0.76}$ & $1.74^{+1.32}_{-0.77}$ & $2.39^{+2.68}_{-0.95}$ & $1.39^{+0.79}_{-0.55}$ & $1.39^{+0.72}_{-0.53}$ & $1.73^{+1.03}_{-0.59}$ & $1.26^{+0.71}_{-0.50}$ & $1.28^{+0.84}_{-0.50}$ & $1.14^{+0.60}_{-0.43}$ & $1.14^{+0.61}_{-0.41}$ & $1.46^{+0.54}_{-0.55}$ \\
$\Delta \Omega ~(\text{deg}^2)$ & $2.67^{+6.43}_{-1.61}$ & $0.53^{+1.40}_{-0.30}$ & $0.59^{+1.30}_{-0.35}$ & $2.01^{+3.91}_{-1.03}$ & $0.12^{+0.23}_{-0.080}$ & $0.11^{+0.20}_{-0.069}$ & $0.11^{+0.18}_{-0.057}$ & $0.054^{+0.088}_{-0.027}$ & $0.053^{+0.091}_{-0.027}$ & $0.029^{+0.048}_{-0.017}$ & $0.028^{+0.050}_{-0.016}$ & $0.036^{+0.055}_{-0.022}$ \\
$\Delta \psi ~(10^{-2}~\rm{rad})$ & $4.99^{+2.92}_{-2.05}$ & $2.48^{+1.23}_{-1.11}$ & $2.43^{+1.26}_{-1.04}$ & $3.57^{+2.77}_{-1.39}$ & $1.71^{+0.92}_{-0.61}$ & $1.68^{+0.86}_{-0.62}$ & $2.10^{+1.25}_{-0.64}$ & $1.57^{+0.82}_{-0.57}$ & $1.66^{+0.89}_{-0.67}$ & $1.38^{+0.70}_{-0.47}$ & $1.39^{+0.72}_{-0.49}$ & $1.81^{+0.76}_{-0.80}$ \\
$\Delta a ~(10^{-6})$ & $23.9^{+16.9}_{-9.67}$ & $11.4^{+7.50}_{-4.40}$ & $11.2^{+7.98}_{-4.56}$ & $19.8^{+11.9}_{-8.24}$ & $9.89^{+7.21}_{-3.79}$ & $9.99^{+5.65}_{-3.59}$ & $14.8^{+8.20}_{-5.67}$ & $9.58^{+5.86}_{-3.72}$ & $9.70^{+6.18}_{-3.65}$ & $8.76^{+5.44}_{-3.28}$ & $8.78^{+5.02}_{-3.15}$ & $9.90^{+5.61}_{-3.24}$ \\
$\Delta e_0 ~(10^{-6})$ & $24.1^{+17.0}_{-9.71}$ & $11.5^{+7.54}_{-4.40}$ & $11.3^{+8.04}_{-4.59}$ & $20.1^{+12.0}_{-8.31}$ & $9.95^{+7.23}_{-3.82}$ & $10.0^{+5.68}_{-3.61}$ & $14.9^{+8.28}_{-5.66}$ & $9.62^{+5.88}_{-3.74}$ & $9.74^{+6.18}_{-3.66}$ & $8.80^{+5.44}_{-3.29}$ & $8.81^{+5.03}_{-3.15}$ & $9.94^{+5.64}_{-3.22}$ \\
$\Delta \Phi_{\phi 0} ~(10^{-2}~\rm{rad})$ & $18.8^{+14.0}_{-8.06}$ & $8.36^{+6.88}_{-3.39}$ & $8.69^{+6.82}_{-3.64}$ & $6.63^{+4.38}_{-2.09}$ & $3.98^{+2.11}_{-1.60}$ & $3.88^{+1.77}_{-1.49}$ & $4.38^{+2.27}_{-1.58}$ & $3.29^{+1.60}_{-1.22}$ & $3.27^{+1.73}_{-1.14}$ & $2.16^{+1.57}_{-0.97}$ & $2.21^{+1.33}_{-0.98}$ & $1.51^{+2.62}_{-0.97}$ \\
$\Delta \Phi_{r 0} ~(10^{-2}~\rm{rad})$ & $8.93^{+6.05}_{-2.70}$ & $4.07^{+2.94}_{-1.21}$ & $4.27^{+2.65}_{-1.43}$ & $4.44^{+2.04}_{-1.62}$ & $2.62^{+1.57}_{-0.92}$ & $2.58^{+1.24}_{-0.81}$ & $3.35^{+1.47}_{-1.11}$ & $2.36^{+1.13}_{-0.76}$ & $2.41^{+1.20}_{-0.78}$ & $2.04^{+1.04}_{-0.69}$ & $2.06^{+0.97}_{-0.66}$ & $3.94^{+2.12}_{-1.10}$ \\
\hline
\end{tabular}
\label{tab:SEMRItable}
\end{sidewaystable*}
\end{widetext}

Another notable feature of Fig.~\ref{fig:SEMRIhist} is that the cumulative distributions for TAIJIp and TAIJIm are nearly indistinguishable across all parameters. This indicates that, for isolated EMRIs, the specific orientation of the TAIJI constellation has a negligible impact on parameter-estimation performance. The rich harmonic structure of the waveforms already provide sufficient information to reduce the impact of orientation-dependent effects.
In contrast, searches for parity violation in the SGWB are much more sensitive to the network geometry. As shown in Ref.~\cite{Chen:2024fto}, the LTm configuration can achieve nearly an order-of-magnitude better sensitivity than LTp owing to its more favorable antenna-pattern correlations.

Taken together, these results illustrate the impact of joint observations on EMRI parameter constraints. A useful comparison can be made between the one-month network observations and the one-year standalone LISA baseline. The three-detector networks operating for only one month not only achieve a larger median SNR than the one-year LISA observation ($114$ versus $101$), but also achieve comparable or superior parameter constraints. In particular, the median sky-localization uncertainty of LTmTQ reaches $0.028~\mathrm{deg}^2$, outperforming the one-year LISA result of $0.036~\mathrm{deg}^2$, while providing a comparable uncertainty of the $\sim9.9\,M_\odot$ precision of the one-year baseline for the primary BH mass. These results indicate that detector networks can achieve parameter constraints comparable to those obtained from substantially longer observations with a single detector. For the EMRI configurations considered here, a one-month observation with a three-detector network yields constraints similar to those obtained from approximately one year of observations with standalone LISA.

\subsection{Resolving concurrent EMRI signals}

In this subsection, we investigate scenarios where two EMRI signals overlap temporally in the data stream. We evaluate two specific configurations: Case 1 involves two secondary compact objects co-orbiting a single primary MBH (assuming the two EMRIs are independent and neglecting multi-body interactions), while Case 2 consists of two separate EMRIs---sharing identical intrinsic parameters ($m_1, m_2, a, e_0$)---originating from different sky directions and orientations.
For both cases, the primary BH has a mass of $m_1 = 10^6~M_{\odot}$ and a dimensionless spin of $a = 0.9$, while the secondary objects are spinless with masses of $m_2 = 10~M_{\odot}$. 
Focusing on Case 1, the target source (Set 1A) is initialized at a fiducial luminosity distance of $d_L = 1~{\rm Gpc}$ with a fixed eccentricity of $e_0 = 0.1$. The companion source (Set 1B) shares the same spatial coordinates ($\lambda, \beta, d_L$), but its eccentricity is uniformly sampled from the interval $[0, 0.5]$. 
To explore the dependence on source parameters, all remaining intrinsic parameters and initial phases are randomly drawn to generate 1000 independent concurrent realizations. Finally, the initial semi-latus rectum $p_0$ for each system is iteratively adjusted to guarantee a consistent observation duration prior to the final plunge.

\begin{figure*}
	\centering
	\includegraphics[width=0.9\textwidth]{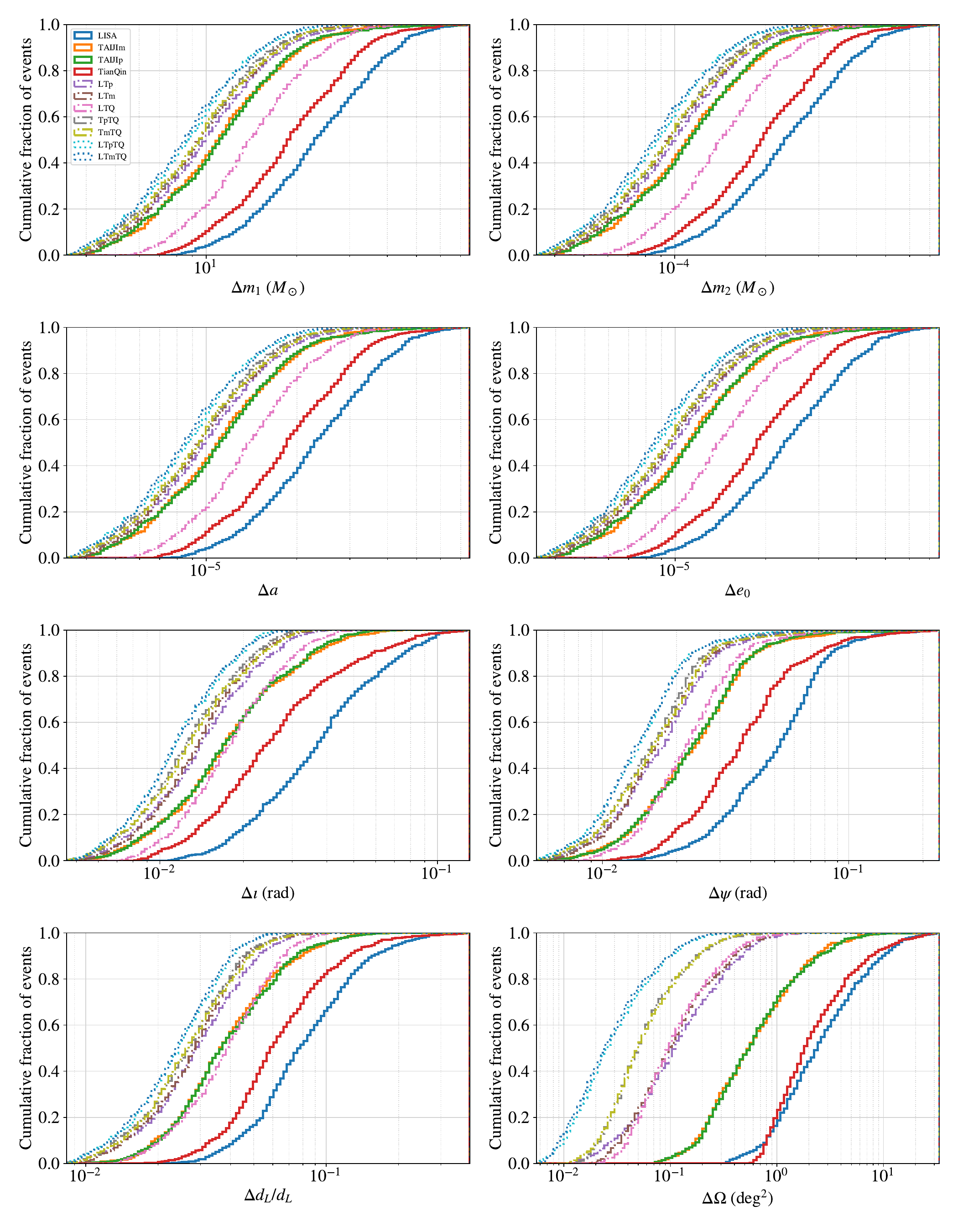}
    \caption{Cumulative distributions of the $1\sigma$ parameter uncertainties for the target EMRI population (Set 1A, with fixed eccentricity $e_0=0.1$) in the concurrent dual-EMRI scenario (Case 1). All results correspond to a one-month observation period. The panels display constraints on intrinsic ($m_1$, $m_2$, $a$, $e_0$) and extrinsic ($\iota$, $\psi$, $d_L$, $\Delta\Omega$) parameters. The systematic leftward shift of the network distributions indicates that the improvements provided by joint observations remain effective in the presence of a concurrent EMRI signal.}
    \label{fig:DEMRIhist_set1A}
\end{figure*}

\begin{figure*}
	\centering
	\includegraphics[width=0.9\textwidth]{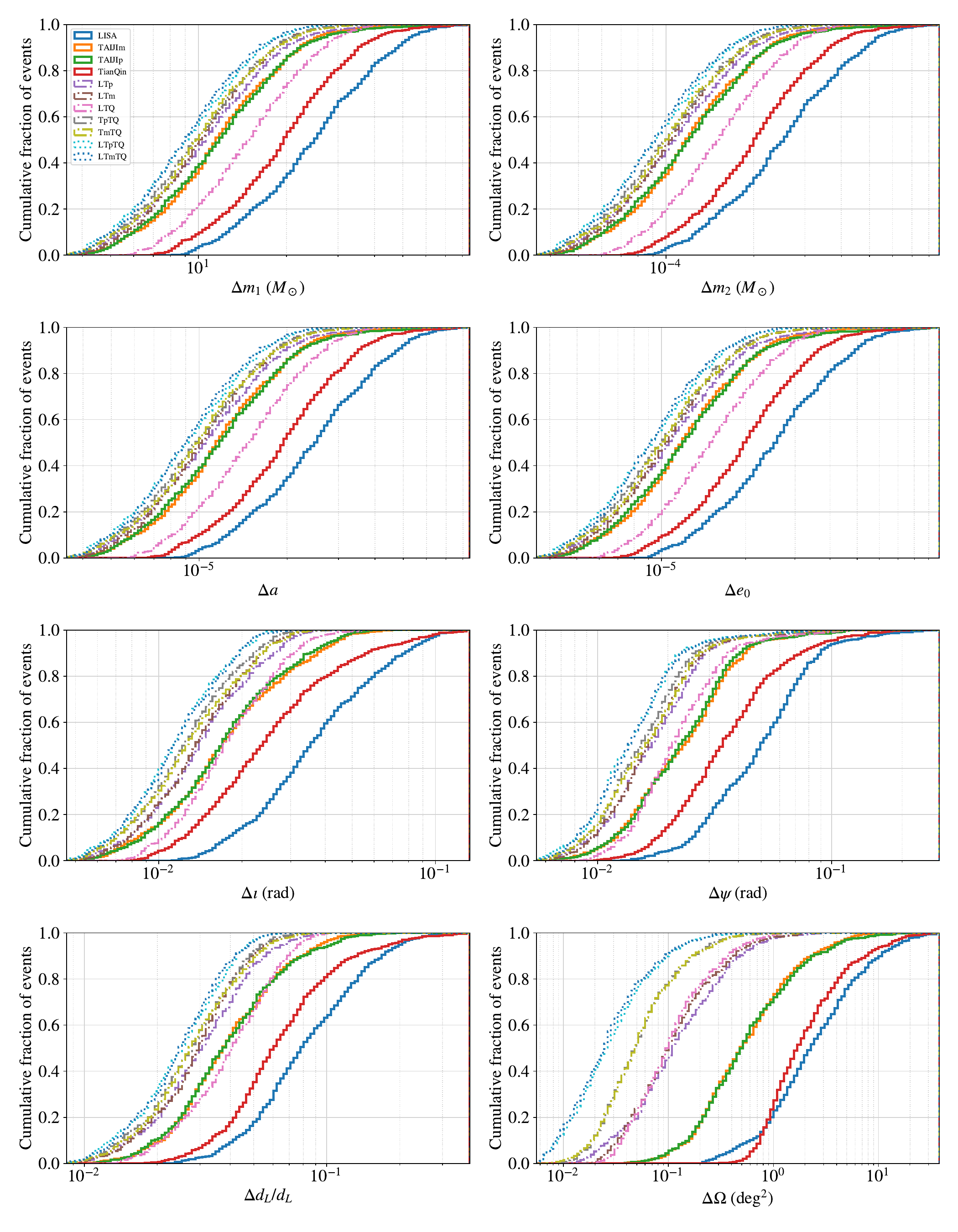}
    \caption{Cumulative distributions of the $1\sigma$ parameter uncertainties for the companion EMRI population (Set 1B), whose initial eccentricities are uniformly sampled in the range $[0,0.5]$, in the concurrent dual-EMRI scenario (Case 1). All results assume a one-month observation period. The trends are broadly consistent with those obtained for Set 1A, indicating that the parameter-estimation performance of the detector networks remains largely unchanged across the range of companion eccentricities considered in this study.}
    \label{fig:DEMRIhist_set1B}
\end{figure*}

Figures~\ref{fig:DEMRIhist_set1A} and \ref{fig:DEMRIhist_set1B}, alongside Table~\ref{tab:dualEMRItable}, detail the Fisher-matrix parameter constraints for this concurrent dual-EMRI scenario (Case 1). As illustrated by the cumulative histograms, the measurement uncertainties for both the target (Set 1A) and companion (Set 1B) populations exhibit the same systematic leftward shift when transitioning from standalone missions to joint networks, consistent with the trend observed for isolated EMRIs. This indicates that the improvements associated with increased network SNR and multiple detector geometries remain present, even in a complex, multi-signal environment.

A direct quantitative comparison between these concurrent results and the isolated EMRI baseline shows little degradation associated with signal overlap. For the target source (Set 1A), the standalone LISA mission yields a median optimal SNR of $42.8$ and a primary mass uncertainty of $\Delta m_1 \approx 23.4~M_\odot$. When the triple-detector network (LTpTQ) is employed, the SNR is increase to $114$, and the $\Delta m_1$ constraint tightens to $8.82~M_\odot$. These constraints are comparable to those obtained in the strictly isolated scenario (Table~\ref{tab:SEMRItable}). 
Similarly, the sky-localization performance remains largely unchanged; the LTmTQ network constrains the Set 1A spatial position to a median precision of $0.025~\text{deg}^2$, outperforming the standalone LISA capability ($2.49~\text{deg}^2$) by two orders of magnitude. This performance indicates that the cross-correlation (off-diagonal) terms between the two signals in the joint FIM are sufficiently small. Consequently, the distinct waveform dynamics and higher-order harmonics lead to weak correlations between the two signals, indicating that source confusion is limited for the configurations considered here in this specific scenario.

The companion source population (Set 1B) exhibits a similar trend, indicating that the parameter constraints are not strongly affected by the variation in the companion's initial eccentricity. For Set 1B, the LTmTQ network yields a median SNR of 110 and a median sky-localization uncertainty of 0.025 deg$^2$, compared with 2.34 deg$^2$ for standalone LISA. The close agreement between the concurrent-source results and the isolated-source baseline suggests that the overlap between the two EMRI signals introduces only weak correlations for the configurations considered here. Consequently, the improvements obtained with the detector networks are primarily associated with the increased SNR and the enhanced geometric information provided by multiple detector baselines, rather than with a substantial additional signal-separation effect.

\begin{sidewaystable*}[htbp]
\centering 
\caption{Median values of the optimal SNR and $1\sigma$ measurement uncertainties for the concurrent dual-EMRI scenario (Case 1), detailing both the target source (Set 1A) and the companion source (Set 1B). The reported values are medians derived from 1000 random source realizations over a one-month observation. Superscripts and subscripts denote the $1\sigma$ asymmetric spread (16th and 84th percentiles) of these constraints across various detector configurations. The constraints exhibit little degradation relative to the isolated-source baseline, indicating that the overlap between the two EMRI signals produces only weak correlations for the source configurations considered here.}
\renewcommand{\arraystretch}{1.5}
\begin{tabular}{l|cccc|ccccc|cc}
\hline
Set 1A & LISA & TAIJIm & TAIJIp & TQ & LTp & LTm & LTQ & TpTQ & TmTQ & LTpTQ & LTmTQ \\
\hline
$\rm{SNR}$ & $42.8^{+27.2}_{-17.1}$ & $90.4^{+59.0}_{-36.5}$ & $88.8^{+68.0}_{-33.7}$ & $52.9^{+34.9}_{-20.8}$ & $99.7^{+69.9}_{-37.7}$ & $104^{+58.9}_{-40.5}$ & $71.2^{+39.2}_{-25.8}$ & $105^{+72.6}_{-37.0}$ & $104^{+67.1}_{-40.7}$ & $114^{+74.1}_{-40.3}$ & $118^{+68.1}_{-46.4}$ \\
$\Delta m_{1}~(M_\odot)$ & $23.4^{+15.6}_{-9.11}$ & $11.1^{+7.51}_{-4.38}$ & $11.3^{+6.90}_{-4.89}$ & $19.1^{+12.2}_{-7.49}$ & $10.0^{+6.11}_{-4.14}$ & $9.68^{+6.19}_{-3.52}$ & $14.1^{+7.97}_{-4.93}$ & $9.61^{+5.20}_{-3.97}$ & $9.58^{+6.11}_{-3.74}$ & $8.82^{+4.77}_{-3.50}$ & $8.56^{+5.50}_{-3.17}$ \\
$\Delta m_{2} ~(10^{-5}~M_\odot)$ & $23.5^{+15.6}_{-9.16}$ & $11.1^{+7.49}_{-4.40}$ & $11.3^{+6.96}_{-4.90}$ & $19.7^{+12.1}_{-7.75}$ & $10.1^{+6.15}_{-4.16}$ & $9.71^{+6.18}_{-3.53}$ & $14.3^{+7.99}_{-5.03}$ & $9.67^{+5.25}_{-3.97}$ & $9.70^{+6.03}_{-3.78}$ & $8.84^{+4.82}_{-3.49}$ & $8.60^{+5.50}_{-3.18}$ \\
$\Delta d_L/d_L ~(10^{-2})$ & $7.82^{+5.94}_{-2.83}$ & $3.69^{+2.81}_{-1.35}$ & $3.74^{+2.84}_{-1.41}$ & $6.09^{+5.28}_{-2.10}$ & $2.92^{+1.84}_{-1.13}$ & $2.87^{+1.41}_{-1.17}$ & $3.92^{+2.25}_{-1.56}$ & $2.74^{+1.47}_{-1.12}$ & $2.69^{+1.75}_{-1.02}$ & $2.46^{+1.32}_{-0.986}$ & $2.42^{+1.28}_{-0.971}$ \\
$\Delta \iota ~(10^{-2}~\rm{rad})$ & $3.72^{+2.94}_{-1.61}$ & $1.73^{+1.41}_{-0.716}$ & $1.72^{+1.38}_{-0.709}$ & $2.49^{+2.59}_{-1.02}$ & $1.41^{+0.823}_{-0.538}$ & $1.39^{+0.677}_{-0.514}$ & $1.81^{+1.04}_{-0.633}$ & $1.26^{+0.750}_{-0.447}$ & $1.28^{+0.843}_{-0.475}$ & $1.15^{+0.616}_{-0.395}$ & $1.15^{+0.600}_{-0.396}$ \\
$\Delta \Omega ~(\text{deg}^2)$ & $2.49^{+5.48}_{-1.49}$ & $0.56^{+1.20}_{-0.34}$ & $0.57^{+1.26}_{-0.35}$ & $1.92^{+4.10}_{-0.98}$ & $0.11^{+0.24}_{-0.071}$ & $0.10^{+0.22}_{-0.059}$ & $0.098^{+0.19}_{-0.051}$ & $0.052^{+0.091}_{-0.026}$ & $0.051^{+0.088}_{-0.025}$ & $0.026^{+0.053}_{-0.014}$ & $0.025^{+0.051}_{-0.014}$ \\
$\Delta \psi~ (10^{-2}~\rm{rad})$ & $5.35^{+2.38}_{-2.39}$ & $2.43^{+1.31}_{-1.02}$ & $2.36^{+1.29}_{-0.964}$ & $3.72^{+2.42}_{-1.54}$ & $1.81^{+0.886}_{-0.705}$ & $1.68^{+0.854}_{-0.563}$ & $2.24^{+1.11}_{-0.762}$ & $1.61^{+0.772}_{-0.571}$ & $1.65^{+0.867}_{-0.624}$ & $1.41^{+0.671}_{-0.478}$ & $1.40^{+0.712}_{-0.471}$ \\
$\Delta a~ (10^{-6})$ & $23.4^{+15.6}_{-9.09}$ & $11.1^{+7.47}_{-4.41}$ & $11.3^{+6.89}_{-4.88}$ & $18.8^{+12.2}_{-7.48}$ & $10.0^{+6.11}_{-4.13}$ & $9.65^{+6.19}_{-3.49}$ & $14.0^{+7.91}_{-4.98}$ & $9.55^{+5.22}_{-3.92}$ & $9.57^{+6.10}_{-3.74}$ & $8.76^{+4.80}_{-3.46}$ & $8.50^{+5.54}_{-3.16}$ \\
$\Delta e_{0} ~(10^{-6})$ & $23.5^{+15.6}_{-9.16}$ & $11.1^{+7.51}_{-4.41}$ & $11.3^{+6.94}_{-4.90}$ & $19.1^{+12.3}_{-7.56}$ & $10.1^{+6.10}_{-4.16}$ & $9.70^{+6.23}_{-3.52}$ & $14.2^{+7.90}_{-5.04}$ & $9.64^{+5.20}_{-3.98}$ & $9.60^{+6.10}_{-3.74}$ & $8.80^{+4.81}_{-3.47}$ & $8.53^{+5.55}_{-3.13}$ \\
$\Delta \Phi_{\phi 0}~ (10^{-2}~\rm{rad})$ & $18.8^{+11.2}_{-8.01}$ & $8.85^{+5.32}_{-3.68}$ & $8.77^{+5.92}_{-3.38}$ & $6.64^{+3.96}_{-2.24}$ & $4.05^{+2.15}_{-1.52}$ & $3.92^{+1.67}_{-1.39}$ & $4.34^{+2.28}_{-1.49}$ & $3.28^{+1.58}_{-1.10}$ & $3.24^{+1.64}_{-1.09}$ & $2.15^{+1.51}_{-0.913}$ & $2.19^{+1.36}_{-0.926}$ \\
$\Delta \Phi_{r 0}~ (10^{-2}~\rm{rad})$ & $8.81^{+5.42}_{-2.51}$ & $4.14^{+2.69}_{-1.23}$ & $4.20^{+2.81}_{-1.28}$ & $4.29^{+2.06}_{-1.45}$ & $2.64^{+1.39}_{-0.991}$ & $2.50^{+1.36}_{-0.777}$ & $3.28^{+1.42}_{-1.01}$ & $2.37^{+1.05}_{-0.823}$ & $2.35^{+1.24}_{-0.786}$ & $2.04^{+0.916}_{-0.727}$ & $2.00^{+1.09}_{-0.644}$ \\
\hline
\hline
Set 1B & LISA & TAIJIm & TAIJIp & TQ & LTp & LTm & LTQ & TpTQ & TmTQ & LTpTQ & LTmTQ \\
\hline
$\rm{SNR}$ & $39.3^{+30.2}_{-16.0}$ & $86.7^{+52.3}_{-36.1}$ & $83.8^{+59.4}_{-33.5}$ & $50.4^{+34.5}_{-19.9}$ & $93.9^{+65.8}_{-36.9}$ & $97.8^{+54.4}_{-38.3}$ & $67.2^{+40.8}_{-25.3}$ & $101^{+64.6}_{-39.0}$ & $100^{+61.9}_{-39.5}$ & $109^{+71.1}_{-42.3}$ & $110^{+62.9}_{-40.9}$ \\
$\Delta m_{1}~(M_\odot)$ & $25.5^{+17.5}_{-11.1}$ & $11.6^{+8.22}_{-4.33}$ & $11.9^{+8.00}_{-4.95}$ & $20.0^{+12.8}_{-8.05}$ & $10.7^{+6.89}_{-4.42}$ & $10.2^{+6.58}_{-3.66}$ & $15.0^{+8.93}_{-5.68}$ & $9.92^{+6.34}_{-3.87}$ & $10.0^{+6.50}_{-3.83}$ & $9.18^{+5.82}_{-3.63}$ & $9.11^{+5.37}_{-3.32}$ \\
$\Delta m_{2} ~(10^{-5}~M_\odot)$ & $25.8^{+17.9}_{-10.9}$ & $11.8^{+8.26}_{-4.50}$ & $12.1^{+8.14}_{-4.87}$ & $20.4^{+12.9}_{-8.09}$ & $10.8^{+6.97}_{-4.40}$ & $10.4^{+6.56}_{-3.75}$ & $15.2^{+9.22}_{-5.67}$ & $10.1^{+6.43}_{-3.99}$ & $10.1^{+6.48}_{-3.83}$ & $9.30^{+5.78}_{-3.65}$ & $9.17^{+5.42}_{-3.28}$ \\
$\Delta d_L/d_L~ (10^{-2})$ & $7.96^{+6.67}_{-3.02}$ & $3.81^{+2.82}_{-1.43}$ & $3.81^{+3.01}_{-1.44}$ & $6.23^{+5.16}_{-2.25}$ & $3.09^{+1.82}_{-1.21}$ & $3.01^{+1.47}_{-1.20}$ & $4.11^{+2.21}_{-1.66}$ & $2.82^{+1.65}_{-1.13}$ & $2.76^{+1.87}_{-1.03}$ & $2.56^{+1.40}_{-1.05}$ & $2.51^{+1.29}_{-0.996}$ \\
$\Delta \iota ~(10^{-2}~\rm{rad})$ & $3.59^{+3.13}_{-1.47}$ & $1.72^{+1.43}_{-0.694}$ & $1.71^{+1.37}_{-0.688}$ & $2.43^{+2.41}_{-1.01}$ & $1.42^{+0.879}_{-0.529}$ & $1.37^{+0.788}_{-0.488}$ & $1.74^{+1.02}_{-0.577}$ & $1.25^{+0.726}_{-0.450}$ & $1.28^{+0.896}_{-0.482}$ & $1.14^{+0.626}_{-0.388}$ & $1.14^{+0.654}_{-0.390}$ \\
$\Delta \Omega ~(\text{deg}^2)$ & $2.34^{+5.61}_{-1.45}$ & $0.51^{+1.15}_{-0.31}$ & $0.51^{+1.23}_{-0.31}$ & $1.77^{+3.45}_{-0.89}$ & $0.11^{+0.26}_{-0.072}$ & $0.10^{+0.23}_{-0.061}$ & $0.10^{+0.17}_{-0.058}$ & $0.050^{+0.080}_{-0.026}$ & $0.050^{+0.080}_{-0.026}$ & $0.026^{+0.049}_{-0.015}$ & $0.025^{+0.047}_{-0.015}$ \\
$\Delta \psi~ (10^{-2}~\rm{rad})$ & $5.04^{+2.91}_{-2.16}$ & $2.37^{+1.40}_{-1.02}$ & $2.27^{+1.34}_{-0.880}$ & $3.48^{+2.78}_{-1.39}$ & $1.72^{+0.944}_{-0.631}$ & $1.66^{+0.853}_{-0.595}$ & $2.10^{+1.22}_{-0.631}$ & $1.56^{+0.824}_{-0.532}$ & $1.62^{+0.869}_{-0.606}$ & $1.38^{+0.669}_{-0.452}$ & $1.37^{+0.726}_{-0.473}$ \\
$\Delta a ~(10^{-6})$ & $25.4^{+17.5}_{-11.1}$ & $11.5^{+8.22}_{-4.33}$ & $11.9^{+7.99}_{-4.94}$ & $19.8^{+12.9}_{-8.03}$ & $10.7^{+6.87}_{-4.40}$ & $10.2^{+6.56}_{-3.65}$ & $14.9^{+8.97}_{-5.61}$ & $9.89^{+6.32}_{-3.86}$ & $9.98^{+6.50}_{-3.82}$ & $9.16^{+5.80}_{-3.61}$ & $9.08^{+5.37}_{-3.31}$ \\
$\Delta e_{0} ~(10^{-6})$ & $26.0^{+18.4}_{-11.1}$ & $12.0^{+8.27}_{-4.68}$ & $12.1^{+8.13}_{-4.99}$ & $20.2^{+12.9}_{-7.97}$ & $10.8^{+7.23}_{-4.42}$ & $10.5^{+6.61}_{-3.93}$ & $15.2^{+9.33}_{-5.74}$ & $10.1^{+6.53}_{-4.01}$ & $10.2^{+6.72}_{-3.96}$ & $9.34^{+5.94}_{-3.70}$ & $9.20^{+5.59}_{-3.30}$ \\
$\Delta \Phi_{\phi 0} ~(10^{-2}~\rm{rad})$ & $16.7^{+12.9}_{-7.29}$ & $7.75^{+4.62}_{-3.24}$ & $7.80^{+5.62}_{-3.39}$ & $6.36^{+4.31}_{-1.80}$ & $4.01^{+2.32}_{-1.52}$ & $3.86^{+2.05}_{-1.43}$ & $4.36^{+2.14}_{-1.50}$ & $3.13^{+1.77}_{-1.17}$ & $3.25^{+1.56}_{-1.19}$ & $2.20^{+1.50}_{-1.02}$ & $2.25^{+1.24}_{-1.03}$ \\
$\Delta \Phi_{r 0} ~(10^{-2}~\rm{rad})$ & $6.67^{+7.44}_{-2.95}$ & $3.07^{+3.42}_{-1.27}$ & $3.09^{+3.56}_{-1.34}$ & $2.42^{+3.66}_{-0.907}$ & $1.57^{+2.10}_{-0.630}$ & $1.51^{+2.00}_{-0.571}$ & $1.73^{+2.81}_{-0.621}$ & $1.29^{+1.82}_{-0.462}$ & $1.31^{+1.92}_{-0.477}$ & $0.979^{+1.64}_{-0.412}$ & $0.984^{+1.77}_{-0.385}$ \\
\hline
\end{tabular}
\label{tab:dualEMRItable}
\end{sidewaystable*}

Table~\ref{tab:dualEMRItable_case2} details the median values of the SNR and the $1\sigma$ measurement uncertainties for the second concurrent scenario (Case 2), which represents another challenging signal-overlap disentanglement. In this configuration, the target source (Set 2A) and the companion source (Set 2B) share completely identical intrinsic parameters ($m_1, m_2, a, e_0$) but possess different extrinsic parameters. Physically, identical intrinsic parameters imply that the two EMRI signals exhibit exactly the same frequency evolution trajectories, rendering them degenerate in the intrinsic frequency space. In this case, source discrimination relies primarily from the distinct spatial projections of the signals onto the detector antenna patterns, driven by their different sky locations and orientations. Despite this overlap in intrinsic waveform evolution, the Fisher-matrix forecasts in Table~\ref{tab:dualEMRItable_case2} demonstrate that the parameter constraints for both populations follow the same systematic tightening observed in the isolated scenarios when transitioning from standalone missions to joint networks.

A quantitative inspection of Set 2A illustrates the network performance in the presence of source overlap.
For a one-month observation, the standalone LISA mission achieves a median optimal SNR of $40.8$ and a primary mass uncertainty of $\Delta m_1 \approx 24.8~M_\odot$. 
When the triple-detector network (LTmTQ) is utilized, the SNR is increase to $114$, and the $\Delta m_1$ constraint is sharply reduced to $8.82~M_\odot$. The sky-localization performance remains comparable to that obtained for isolated sources; the LTmTQ network constrains the spatial position of Set 2A to $0.026~\text{deg}^2$, which is two orders of magnitude tighter than the standalone LISA capability ($2.85~\text{deg}^2$). 
The companion source (Set 2B) exhibits identical performance metrics, with the LTmTQ network achieving a sky localization precision of $0.027~\text{deg}^2$. 

The close agreement between the concurrent-source results and the isolated-source baseline suggests that the correlations between the two signals remain weak for the configurations considered here. Even when the two EMRIs share identical intrinsic parameters, differences in sky location and orientation lead to distinct detector responses, reducing the impact of signal overlap on the Fisher-matrix constraints. Since similar behavior is observed for both standalone and network observations, the primary effect of the detector networks remains the same as in the isolated-source case: increased SNR and improved sky localization arising from the combination of multiple detector baselines and antenna patterns. As a result, the parameter constraints obtained for concurrent EMRIs remain comparable to those obtained for isolated sources throughout the scenarios examined in this work.

\begin{sidewaystable*}[htbp]
\centering 
\caption{Median values of the optimal SNR and the corresponding $1\sigma$ parameter uncertainties for the concurrent dual-EMRI scenario (Case 2). In this configuration, the target source (Set 2A) and companion source (Set 2B) share identical intrinsic parameters ($m_1$, $m_2$, $a$, $e_0$) but differ in their extrinsic parameters. The quoted medians and uncertainty ranges correspond to the 50th, 16th, and 84th percentiles obtained from 1000 realizations with a one-month observation period. The resulting parameter constraints remain comparable to those obtained for isolated EMRIs, indicating that no substantial degradation is observed within the parameter space explored in this study.}
\renewcommand{\arraystretch}{1.5}
\begin{tabular}{l|cccc|ccccc|cc}
\hline
Set 2A & LISA & TAIJIm & TAIJIp & TQ & LTp & LTm & LTQ & TpTQ & TmTQ & LTpTQ & LTmTQ \\
\hline
$\rm{SNR}$ & $40.8^{+28.4}_{-16.2}$ & $87.1^{+60.7}_{-33.9}$ & $87.8^{+59.4}_{-36.4}$ & $51.3^{+34.2}_{-19.8}$ & $99.9^{+60.0}_{-41.7}$ & $101^{+61.7}_{-38.5}$ & $68.3^{+38.4}_{-25.1}$ & $104^{+63.7}_{-40.9}$ & $102^{+70.8}_{-38.8}$ & $114^{+64.8}_{-44.5}$ & $113^{+71.4}_{-43.3}$ \\
$\Delta m_{1}~M_\odot$ & $24.8^{+16.0}_{-10.1}$ & $11.7^{+7.21}_{-4.89}$ & $11.5^{+8.15}_{-4.58}$ & $20.0^{+12.3}_{-7.39}$ & $10.1^{+7.12}_{-3.88}$ & $10.0^{+6.16}_{-3.81}$ & $14.8^{+8.40}_{-5.09}$ & $9.64^{+6.24}_{-3.62}$ & $9.91^{+6.07}_{-4.09}$ & $8.82^{+5.63}_{-3.24}$ & $8.87^{+5.43}_{-3.08}$ \\
$\Delta m_{2}~(10^{-5})~M_\odot$ & $25.3^{+16.4}_{-10.2}$ & $12.0^{+7.37}_{-5.07}$ & $11.8^{+8.15}_{-4.59}$ & $20.7^{+12.7}_{-7.62}$ & $10.3^{+7.08}_{-3.81}$ & $10.2^{+6.20}_{-3.87}$ & $15.0^{+8.64}_{-5.32}$ & $9.81^{+6.27}_{-3.59}$ & $10.1^{+6.17}_{-4.15}$ & $8.95^{+5.63}_{-3.27}$ & $9.02^{+5.53}_{-3.09}$ \\
$\Delta d_L/d_L ~(10^{-2})$ & $8.38^{+5.96}_{-3.03}$ & $4.00^{+2.71}_{-1.51}$ & $3.86^{+2.98}_{-1.30}$ & $6.47^{+5.16}_{-2.37}$ & $3.01^{+1.86}_{-1.20}$ & $3.02^{+1.30}_{-1.26}$ & $4.05^{+2.09}_{-1.52}$ & $2.73^{+1.66}_{-0.956}$ & $2.84^{+1.64}_{-1.17}$ & $2.48^{+1.45}_{-0.954}$ & $2.53^{+1.19}_{-0.964}$ \\
$\Delta \iota ~(10^{-2}~\rm{rad})$ & $3.76^{+2.94}_{-1.61}$ & $1.80^{+1.45}_{-0.784}$ & $1.79^{+1.30}_{-0.764}$ & $2.59^{+2.40}_{-1.11}$ & $1.45^{+0.767}_{-0.534}$ & $1.37^{+0.785}_{-0.482}$ & $1.79^{+0.986}_{-0.596}$ & $1.27^{+0.720}_{-0.458}$ & $1.29^{+0.840}_{-0.489}$ & $1.15^{+0.590}_{-0.409}$ & $1.14^{+0.615}_{-0.390}$ \\
$\Delta \Omega ~(\text{deg}^2)$ & $2.85^{+6.47}_{-1.72}$ & $0.60^{+1.34}_{-0.36}$ & $0.59^{+1.48}_{-0.35}$ & $1.96^{+3.84}_{-0.96}$ & $0.11^{+0.24}_{-0.062}$ & $0.11^{+0.19}_{-0.064}$ & $0.10^{+0.15}_{-0.055}$ & $0.054^{+0.080}_{-0.027}$ & $0.054^{+0.079}_{-0.027}$ & $0.027^{+0.044}_{-0.015}$ & $0.026^{+0.043}_{-0.015}$ \\
$\Delta \psi ~(10^{-2}~\rm{rad})$ & $5.27^{+3.10}_{-2.18}$ & $2.52^{+1.39}_{-1.07}$ & $2.54^{+1.40}_{-1.13}$ & $3.78^{+2.42}_{-1.55}$ & $1.71^{+0.979}_{-0.621}$ & $1.69^{+0.847}_{-0.581}$ & $2.20^{+1.12}_{-0.692}$ & $1.59^{+0.866}_{-0.556}$ & $1.70^{+0.789}_{-0.670}$ & $1.38^{+0.752}_{-0.451}$ & $1.40^{+0.659}_{-0.475}$ \\
$\Delta a~ (10^{-6})$ & $24.8^{+16.0}_{-10.1}$ & $11.7^{+7.21}_{-4.88}$ & $11.5^{+8.10}_{-4.59}$ & $19.7^{+12.1}_{-7.91}$ & $10.1^{+7.12}_{-3.78}$ & $9.99^{+6.14}_{-3.81}$ & $14.7^{+8.42}_{-5.25}$ & $9.61^{+6.19}_{-3.61}$ & $9.89^{+6.07}_{-4.08}$ & $8.81^{+5.61}_{-3.20}$ & $8.86^{+5.42}_{-3.45}$ \\
$\Delta e_{0}~ (10^{-6})$ & $25.5^{+16.8}_{-10.1}$ & $12.2^{+7.55}_{-5.12}$ & $12.0^{+8.11}_{-4.72}$ & $20.6^{+12.7}_{-8.17}$ & $10.4^{+7.08}_{-3.89}$ & $10.2^{+6.30}_{-3.83}$ & $14.9^{+8.67}_{-5.21}$ & $9.88^{+6.15}_{-3.73}$ & $10.1^{+6.25}_{-4.11}$ & $8.97^{+5.58}_{-3.30}$ & $8.97^{+5.57}_{-3.43}$ \\
$\Delta \Phi_{\phi 0} ~(10^{-2}~\rm{rad})$ & $19.6^{+15.4}_{-8.37}$ & $9.05^{+6.34}_{-3.39}$ & $9.08^{+7.35}_{-3.72}$ & $6.88^{+4.45}_{-2.13}$ & $4.06^{+2.23}_{-1.42}$ & $4.10^{+1.65}_{-1.50}$ & $4.47^{+2.16}_{-1.33}$ & $3.38^{+1.66}_{-1.08}$ & $3.49^{+1.45}_{-1.20}$ & $2.27^{+1.47}_{-0.984}$ & $2.34^{+1.13}_{-1.03}$ \\
$\Delta \Phi_{r 0}~ (10^{-2}~\rm{rad})$ & $9.25^{+6.43}_{-2.89}$ & $4.34^{+2.89}_{-1.35}$ & $4.27^{+2.93}_{-1.31}$ & $4.43^{+2.09}_{-1.51}$ & $2.65^{+1.55}_{-0.938}$ & $2.59^{+1.40}_{-0.842}$ & $3.32^{+1.55}_{-1.00}$ & $2.37^{+1.22}_{-0.702}$ & $2.44^{+1.16}_{-0.857}$ & $2.04^{+1.11}_{-0.657}$ & $2.06^{+1.06}_{-0.696}$ \\
\hline
\hline
Set 2B & LISA & TAIJIm & TAIJIp & TQ & LTp & LTm & LTQ & TpTQ & TmTQ & LTpTQ & LTmTQ \\
\hline
$\rm{SNR}$ & $42.4^{+27.6}_{-18.3}$ & $90.2^{+57.5}_{-37.4}$ & $90.2^{+59.7}_{-36.3}$ & $52.5^{+34.1}_{-20.8}$ & $101^{+63.2}_{-39.4}$ & $103^{+58.3}_{-39.6}$ & $69.2^{+38.4}_{-25.6}$ & $106^{+68.6}_{-40.4}$ & $106^{+65.6}_{-42.7}$ & $116^{+69.1}_{-45.1}$ & $117^{+65.8}_{-44.7}$ \\
$\Delta m_{1}~M_\odot$ & $24.0^{+17.8}_{-9.57}$ & $11.3^{+7.87}_{-4.40}$ & $11.2^{+7.45}_{-4.54}$ & $19.4^{+12.6}_{-7.39}$ & $10.0^{+6.53}_{-3.88}$ & $9.76^{+6.08}_{-3.52}$ & $14.5^{+8.64}_{-5.09}$ & $9.45^{+5.87}_{-3.63}$ & $9.49^{+6.35}_{-3.63}$ & $8.66^{+5.45}_{-3.24}$ & $8.56^{+5.23}_{-3.08}$ \\
$\Delta m_{2}~ (10^{-5}~M_\odot)$ & $24.6^{+17.5}_{-9.75}$ & $11.5^{+7.92}_{-4.42}$ & $11.5^{+7.60}_{-4.53}$ & $20.2^{+12.6}_{-7.62}$ & $10.2^{+6.47}_{-3.90}$ & $9.90^{+6.16}_{-3.53}$ & $14.9^{+8.65}_{-5.18}$ & $9.60^{+5.89}_{-3.67}$ & $9.71^{+6.31}_{-3.71}$ & $8.77^{+5.51}_{-3.26}$ & $8.66^{+5.24}_{-3.09}$ \\
$\Delta d_L/d_L ~(10^{-2})$ & $8.23^{+6.36}_{-3.13}$ & $3.91^{+3.12}_{-1.51}$ & $3.77^{+2.52}_{-1.34}$ & $6.26^{+5.60}_{-2.07}$ & $3.01^{+1.65}_{-1.20}$ & $2.93^{+1.42}_{-1.16}$ & $3.95^{+2.28}_{-1.49}$ & $2.74^{+1.48}_{-1.09}$ & $2.70^{+1.88}_{-0.997}$ & $2.44^{+1.31}_{-0.954}$ & $2.45^{+1.33}_{-0.964}$ \\
$\Delta \iota~ (10^{-2}~\rm{rad})$ & $3.75^{+2.72}_{-1.53}$ & $1.77^{+1.39}_{-0.759}$ & $1.72^{+1.13}_{-0.687}$ & $2.53^{+2.11}_{-1.03}$ & $1.40^{+0.634}_{-0.534}$ & $1.40^{+0.683}_{-0.507}$ & $1.77^{+0.714}_{-0.590}$ & $1.26^{+0.695}_{-0.471}$ & $1.30^{+0.775}_{-0.492}$ & $1.14^{+0.578}_{-0.409}$ & $1.15^{+0.572}_{-0.392}$ \\
$\Delta \Omega ~(\text{deg}^2)$ & $2.77^{+3.51}_{-1.60}$ & $0.65^{+1.42}_{-0.39}$ & $0.62^{+1.54}_{-0.37}$ & $2.05^{+3.51}_{-1.01}$ & $0.12^{+0.21}_{-0.077}$ & $0.11^{+0.20}_{-0.066}$ & $0.11^{+0.19}_{-0.058}$ & $0.055^{+0.078}_{-0.029}$ & $0.056^{+0.066}_{-0.028}$ & $0.028^{+0.048}_{-0.017}$ & $0.027^{+0.047}_{-0.016}$ \\
$\Delta \psi~ (10^{-2}~\rm{rad})$ & $5.32^{+2.78}_{-2.33}$ & $2.54^{+1.38}_{-1.08}$ & $2.47^{+1.36}_{-1.05}$ & $3.84^{+2.20}_{-1.62}$ & $1.73^{+0.615}_{-0.621}$ & $1.74^{+0.605}_{-0.595}$ & $2.21^{+0.933}_{-0.714}$ & $1.57^{+0.694}_{-0.541}$ & $1.62^{+0.736}_{-0.596}$ & $1.36^{+0.616}_{-0.432}$ & $1.41^{+0.591}_{-0.479}$ \\
$\Delta a~ (10^{-6})$ & $24.0^{+17.8}_{-9.56}$ & $11.3^{+7.87}_{-4.39}$ & $11.2^{+7.46}_{-4.52}$ & $19.1^{+12.5}_{-7.37}$ & $10.0^{+6.47}_{-3.87}$ & $9.75^{+6.08}_{-3.52}$ & $14.5^{+8.59}_{-5.12}$ & $9.44^{+5.84}_{-3.64}$ & $9.50^{+6.31}_{-3.63}$ & $8.64^{+5.45}_{-3.23}$ & $8.55^{+5.22}_{-3.09}$ \\
$\Delta e_{0}~ (10^{-6})$ & $25.0^{+18.4}_{-10.0}$ & $11.6^{+7.90}_{-4.44}$ & $11.6^{+7.63}_{-4.43}$ & $19.8^{+12.8}_{-7.58}$ & $10.2^{+6.62}_{-3.96}$ & $10.0^{+6.14}_{-3.68}$ & $14.7^{+8.76}_{-5.08}$ & $9.66^{+5.84}_{-3.71}$ & $9.64^{+6.31}_{-3.68}$ & $8.84^{+5.51}_{-3.30}$ & $8.73^{+5.01}_{-3.16}$ \\
$\Delta \Phi_{\phi 0}~ (10^{-2}~\rm{rad})$ & $19.4^{+13.3}_{-7.65}$ & $9.02^{+6.14}_{-3.51}$ & $9.20^{+6.66}_{-3.68}$ & $6.69^{+3.45}_{-2.00}$ & $3.98^{+1.39}_{-1.39}$ & $3.97^{+1.44}_{-1.36}$ & $4.37^{+1.36}_{-1.32}$ & $3.27^{+1.08}_{-1.12}$ & $3.34^{+1.23}_{-1.16}$ & $2.16^{+0.867}_{-0.943}$ & $2.23^{+0.943}_{-0.943}$ \\
$\Delta \Phi_{r 0} ~(10^{-2}~\rm{rad})$ & $9.12^{+9.12}_{-2.63}$ & $4.29^{+2.80}_{-1.31}$ & $4.23^{+2.91}_{-1.24}$ & $4.33^{+2.02}_{-1.47}$ & $2.64^{+1.11}_{-0.956}$ & $2.58^{+1.06}_{-0.808}$ & $3.27^{+1.03}_{-0.980}$ & $2.35^{+0.751}_{-0.647}$ & $2.35^{+0.736}_{-0.681}$ & $2.03^{+0.678}_{-0.657}$ & $2.02^{+0.627}_{-0.627}$ \\
\hline
\end{tabular}
\label{tab:dualEMRItable_case2}
\end{sidewaystable*}

\section{Conclusion and discussions} \label{sec6}

In this work, we investigated the capability of space-based GW detector networks composed of LISA, TAIJI, and TianQin to constrain the parameters of both isolated and concurrent EMRIs. Using the FIM formalism together with EMRI waveforms generated by the \texttt{FEW} framework, we evaluated the expected measurement uncertainties for a variety of dual- and triple-detector network configurations. Particular attention was paid to the extent to which joint observations can improve parameter constraints during a relatively short one-month observation period.

A principal result of this study is that joint observations substantially enhance EMRI parameter constraints relative to a standalone detector.
For the intrinsic parameters, including the masses $(m_1,m_2)$ and the dimensionless spin $a$, the improvement closely follows the expected $1/\mathrm{SNR}$ scaling, indicating that the dominant contribution of the detector networks arises from the increase in accumulated signal power.
Within the source population considered here, the three-detector networks achieve parameter constraints after one month that are comparable to those obtained from a one-year observation by LISA alone. 
This result suggests that detector networks can significantly reduce the observation time required to reach a given level of measurement precision.
The improvement is particularly pronounced for sky localization. In contrast to the intrinsic parameters, localization benefits not only from the increased SNR but also from the geometric diversity provided by multiple constellations. The large spatial separations and distinct antenna-pattern responses of the detectors introduce additional timing and phase information, which helps break sky-position degeneracies present in single-detector observations. As a consequence, the median sky-localization uncertainty can be reduced by up to two orders of magnitude for the network configurations considered in this work.

We also investigated two concurrent-EMRI scenarios in order to assess the impact of overlapping signals on parameter constraints. In both cases, the inferred uncertainties remain close to those obtained for isolated sources. This behavior indicates that the cross-correlation terms between the two signals are generally small for the source configurations considered here. 
Even when the two EMRIs share identical intrinsic parameters, differences in sky position and orientation lead to distinct detector responses, reducing the level of parameter correlation.
Consequently, the network configurations retain most of their parameter-constraining capability in the presence of concurrent signals.

Our results further indicate that the choice between the TAIJIp and TAIJIm orbital configurations has only a minor impact on EMRI parameter constraints. For the EMRI population considered here, the cumulative SNR and parameter uncertainties obtained with the two configurations are nearly indistinguishable. This contrasts with previous studies of other science cases, such as SGWB polarization measurements and MBBH localization, where the relative orientation between detector constellations plays a more significant role~\cite{Wang:2021uih,Chen:2024fto}. Therefore, while the choice of TAIJI orbit may be important for broader network science objectives, its influence on EMRI parameter constraints appears limited within the framework adopted in this work.

Several limitations of the present analysis should be noted. The FIM formalism provides an estimate of the local parameter uncertainties under the assumptions of Gaussian noise and sufficiently high SNR. As a result, it does not capture potential non-Gaussian features of the likelihood surface, nor does it directly address the practical challenges associated with extracting multiple overlapping EMRI signals from realistic detector data. In particular, the high dimensionality of the EMRI parameter space and the computational cost of waveform generation remain significant obstacles for full Bayesian analyses.

Nevertheless, the results presented here provide a useful quantitative assessment of the information content available in future space-based detector networks. They indicate that the combination of multiple constellations can substantially improve EMRI parameter constraints, especially for source localization, while maintaining comparable performance in the presence of overlapping signals. Future studies incorporating more general EMRI orbital configurations, environmental effects, and full Bayesian inference methods will be important for further evaluating the scientific capabilities of global space-based GW detector networks.

\begin{acknowledgments}
C.Z. was supported by the National Natural Science Foundation of China under Grant No. 12505076.
G.W. was supported by the National Key Research and Development Program of China under Grant No. 2021YFC2201903 and NSFC Grant No. 12575058.
\end{acknowledgments}

\bibliographystyle{apsrev4-2f}
\bibliography{ref}

\end{document}